\documentclass[twocolumn,twocolappendix]{aastex631}

\usepackage{apjfonts}
\usepackage{graphicx}
\usepackage{wrapfig}
\usepackage{amsmath}
\usepackage{multirow}
\usepackage{booktabs}
\usepackage{float}
\usepackage{natbib}
\bibpunct[; ]{(}{)}{,}{a}{}{,}
\newcommand{\Msun}{$M_{\sun}$}

\newcommand{\solperyr}{$M_{\sun}$ yr$^{-1}$}

\newcommand{\SigSFR}{$\Sigma_{\mathrm{SFR}}$}
\newcommand{\CSigSFR}{$\langle \Sigma_{\mathrm{SFR}} \rangle$}

\newcommand{\um}{$\mu$m}

\defcitealias{Johnson15_AP}{J15}

\received{January 21, 2022}
\revised{August 14, 2022}
\accepted{August 21, 2022}
\submitjournal{ApJ}
\shorttitle{PHATTER. IV. Star Cluster Catalog}
\shortauthors{Johnson et al.}

\begin{document}

\title{The Panchromatic Hubble Andromeda Treasury: Triangulum Extended Region (PHATTER).\\
IV. Star Cluster Catalog}

\author[0000-0001-6421-0953]{L. Clifton Johnson}
\affiliation{Center for Interdisciplinary Exploration and Research in Astrophysics (CIERA) and Department of Physics and Astronomy, Northwestern University, 1800 Sherman Ave., Evanston, IL 60201, USA}

\author[0000-0001-6320-2230]{Tobin M. Wainer}
\affiliation{Department of Physics and Astronomy, University of Utah, Salt Lake City, UT 84112, USA}
\affiliation{Center for Interdisciplinary Exploration and Research in Astrophysics (CIERA) and Department of Physics and Astronomy, Northwestern University, 1800 Sherman Ave., Evanston, IL 60201, USA}

\author[0000-0001-9961-8203]{Estephani E. TorresVillanueva}
\affiliation{Department of Physics and Astronomy, University of Utah, Salt Lake City, UT 84112, USA}
\affiliation{Department of Astronomy, University of Wisconsin-Madison, Madison, WI, 53706, USA}

\author[0000-0003-0248-5470]{Anil C. Seth}
\affiliation{Department of Physics and Astronomy, University of Utah, Salt Lake City, UT 84112, USA}

\author[0000-0002-7502-0597]{Benjamin F. Williams}
\affiliation{Department of Astronomy, University of Washington, Box 351580, Seattle, WA 98195, USA}

\author[0000-0001-7531-9815]{Meredith J. Durbin}
\affiliation{Department of Astronomy, University of Washington, Box 351580, Seattle, WA 98195, USA}

\author[0000-0002-1264-2006]{Julianne J. Dalcanton}
\affiliation{Department of Astronomy, University of Washington, Box 351580, Seattle, WA 98195, USA}
\affiliation{Center for Computational Astrophysics, Flatiron Institute, 162 Fifth Avenue, New York, NY 10010, USA}

\author[0000-0002-6442-6030]{Daniel R. Weisz}
\affiliation{Department of Astronomy, University of California, Berkeley, CA 94720, USA}

\author[0000-0002-5564-9873]{Eric F. Bell}
\affiliation{Department of Astronomy, University of Michigan, 323 West Hall, 1085 S. University Ave., Ann Arbor, MI 48105, USA}

\author[0000-0001-8867-4234]{Puragra Guhathakurta}
\affiliation{University of California - Santa Cruz, 1156 High Street, Santa Cruz, CA 95064, USA}

\author[0000-0003-0605-8732]{Evan Skillman}
\affiliation{Minnesota Institute for Astrophysics, 116 Church Street SE, Minneapolis, MN 55455, USA}

\author[0000-0003-2599-7524]{Adam Smercina}
\affiliation{Department of Astronomy, University of Washington, Box 351580, Seattle, WA 98195, USA}

\author{PHATTER Collaboration}
\noaffiliation

\correspondingauthor{L. Clifton Johnson}
\email{lcj@northwestern.edu}

\begin{abstract}
We construct a catalog of star clusters from Hubble Space Telescope images of the inner disk of the Triangulum Galaxy (M33) using image classifications collected by the Local Group Cluster Search, a citizen science project hosted on the Zooniverse platform. We identify 1214 star clusters within the Hubble Space Telescope imaging footprint of the Panchromatic Hubble Andromeda Treasury: Triangulum Extended Region (PHATTER) survey.  Comparing this catalog to existing compilations in the literature, 68\% of the clusters are newly identified.  The final catalog includes multi-band aperture photometry and fits for cluster properties via integrated light SED fitting. The cluster catalog's 50\% completeness limit is $\sim$1500 \Msun\ at an age of 100 Myr, as derived from comprehensive synthetic cluster tests.
\end{abstract}



\section{Introduction} \label{sec:intro}

For decades, star clusters have been recognized as useful tracers of star formation. Rather than representing uniform tracers of star formation, we now understand that non-embedded, long-lived, gravitationally-bound star clusters emerge from natal regions with relatively high gas surface density and star formation efficiency \citep{Elmegreen97, Kruijssen12, Grudic21}. Star clusters contain $\sim$5--30\% of the stellar mass formed \citep{Johnson16_gamma, Adamo20} and are long-lasting remnants of peaks in hierarchically structured star-forming regions that survived the stellar feedback and gas removal dissolution that unbind most stellar groupings and natal structures.

We now have broad samples of clusters spanning a wide variety of galactic environments, where long-lived bound star clusters are tell-tale tracers of past episodes of intense, efficient star formation. Large imaging surveys using the \textit{Hubble Space Telescope} (HST) have made significant progress in cataloging and characterizing star cluster populations in nearby (3--30 Mpc) galaxies \citep[e.g., LEGUS, PHANGS-HST;][]{Adamo17, Lee22}.  While the diversity of galactic environments included in these samples is very useful for purposes of galaxy-to-galaxy comparisons, individual star clusters at these distances are only marginally resolved, limiting observational measurements to integrated properties.

In contrast, studies of neighboring galaxies in the Local Group provide a unique opportunity for detailed studies of external galaxies and their star clusters, yielding a rich picture of star formation observed at and below molecular cloud spatial scales --- a level of detail not possible in more distant extragalactic targets.  Due to their proximity and the spatial resolving power of HST, Local Group galaxies provide an unmatched opportunity to construct high quality cluster catalogs and make detailed observations of these systems and their environments \citep[e.g.,][]{Johnson12}. Local Group cluster catalogs reach low cluster mass completeness limits leading to increased sample sizes and diversity. Observations at these distances resolve individual cluster member stars leading to marked improvements in age-dating precision and usefulness to stellar evolution studies \citep{Johnson16_gamma, Girardi20}.  

This paper studies the star cluster population of the Triangulum Galaxy (M33), whose intermediate galaxy mass and relatively active star formation providing a point of comparison to studies of the stellar cluster populations of the Andromeda Galaxy (M31) and the Magellanic Clouds.  Notably, M33 hosts a larger star formation rate surface density (\SigSFR) than the bulge-dominated, relatively quiescent M31 \citep{Williams21}. Therefore, we expect observations of M33's young cluster population to unlock valuable new insight into star cluster formation and evolution. Triangulum's relatively face-on orientation \citep[inclination angle of 55$^{\circ}$;][]{Koch18} also presents an advantage over Andromeda in terms of line-of-sight dust attenuation and projection effects.

Studies of the star cluster population in M33 include results from both ground-based \citep[e.g.,][]{Christian82, SanRoman10} and space-based \citep[e.g.,][]{Chandar99, Chandar01, Park07, SanRoman09} observations, as summarized by \citet{Sarajedini07}. Previous work demonstrated HST's utility for identifying star clusters, but spatial coverage of M33's star-forming disk was sparse and largely non-contiguous, preventing systematic studies of the cluster population. As a result, much of the past work to characterize M33's star clusters makes use of ground-based imaging and photometry \citep[e.g.,][]{deMeulenaer15, Fan14}, especially from the Local Group Galaxy Survey \citep[LGGS;][]{Massey06}.

We note that an alternative catalog of M33 young star cluster candidates was published by \citet{Sharma11} based on mid-infrared Spitzer 24\um\ source identification.  This catalog should be sensitive to embedded clusters that are not detected by an optical search, and has been used for analysis of M33 clusters by a number of groups \citep[e.g.,][]{GonzalezLopezlira12, Pflamm13, Corbelli17}.  However, significant concerns about this catalog's contamination by non-cluster objects and its suitability for star cluster studies have been well articulated by \citet{Sun16}. HST observations will have comparatively limited sensitivity to the earliest embedded stages of star cluster formation (1--3 Myr), but its high spatial resolution (0.1 arcsec v.\ $\sim$6 arcsec for Spitzer 24$\mu$m images) remains the best avenue for identification and analysis of star clusters at nearly every other age.

The Panchromatic Hubble Andromeda Treasury: Triangulum Extended Region survey \citep[PHATTER;][]{Williams21} of M33 delivers contiguous, multi-band imaging of a majority of the galaxy's star-forming disk, extending the same quality of data obtained by the Panchromatic Hubble Andromeda Treasury survey \citep[PHAT;][]{Dalcanton12} in M31 to M33.  Similarly, this work moves cluster studies in M33 into a new era using techniques and analysis that were employed to construct the PHAT cluster catalog \citep[][hereafter J15]{Johnson15_AP} for M31.

Facing the absence of a robust algorithmic method for identifying clusters in Local Group galaxy images, we launched an online citizen science project, the Local Group Cluster Search (LGCS), to perform a visual search of the PHATTER data. We employ the crowdsourced methodology developed for PHAT and the Andromeda Project \citepalias{Johnson15_AP} to construct a star cluster catalog.  This approach improves on the subjectivity of expert-led searches conducted in the past in M33 \citep[e.g.,][]{Christian82, Chandar99} using a ``wisdom of the crowds'' consensus classification technique, where an unbiased, repeatable result is obtained by averaging over tens of independent image classifications.  In addition to producing a robust cluster catalog, we characterize catalog completeness using synthetic clusters inserted into the search images. Not only does this methodology produce useful results, but it facilitates meaningful engagement with project volunteers regarding astronomy and star cluster science.

In this paper, we present the survey-wide cluster catalog for PHATTER. We describe the LGCS project, its input data and preparation, and data collection results in \S\ref{sec:data}.  We analyze image classifications and outline the steps required to produce a cluster catalog in \S\ref{sec:catcon}.  We present the final catalog in \S\ref{sec:catalog}, followed by a characterization of the catalog's completeness in \S\ref{sec:catcompleteness}. We derive integrated light ages and masses in \S\ref{sec:slugfitting} and place the new catalog in context with previous work in M33 and similar work in M31 in \S\ref{sec:discussion}.

This catalog serves as the foundation for PHATTER survey cluster science.  Future work includes the measurement of the cluster mass function \citep{Wainer22}, measurement of the high-mass stellar initial mass function, calibration of stellar evolution models, and more.  These studies will build upon and benefit from comparisons to the PHAT star cluster studies of M31, including measurements of star cluster formation efficiency \citep{Johnson16_gamma}, the cluster mass function \citep{Johnson17}, and the high mass stellar initial mass function \citep{Weisz15}.

Throughout this work, we assume a distance to M33 of 859 kpc \citep[distance modulus: 24.67][]{deGrijs14} where 1 arcsec is equivalent to $\sim$4.2 pc.

\section{Data} \label{sec:data}

In this section, we describe the Local Group Cluster Search citizen science project and the underlying HST data that enables this study. We begin by describing the PHATTER imaging used for the project (\S\ref{sec:phatdata}) and the LGCS website interface (\S\ref{sec:interface}).  Next, we discuss data collection and statistics regarding image classifications and project volunteers (\S\ref{sec:datacollection}). Finally, we discuss the creation of synthetic clusters used to characterize catalog completeness (\S\ref{sec:synclst}).

\subsection{PHATTER Images and Resolved Star Photometry} \label{sec:phatdata}

The HST images analyzed by the LGCS project were obtained as part of the PHATTER survey.  Full details of the survey are presented in \citet{Williams21}, but here we highlight the features of this survey that are relevant to star cluster catalog work.

The PHATTER survey uses the same imaging strategy as the PHAT survey, where parallel observations are efficiently obtained with the Advanced Camera for Surveys (ACS) and Wide Field Camera 3 (WFC3). These observations are organized into three contiguous ``bricks'', a 3$\times$6 mosaic of WFC3 footprints formed from pairs of parallel ACS and WFC3 images that combine to create a rectangular region of fully-overlapped spatial coverage in all observed passbands.  This observing strategy yields images in six filters: F475W and F814W in the optical obtained with ACS/WFC; F275W and F336W in the near-UV obtained with WFC3/UVIS; F110W and F160W in the NIR obtained with WFC3/IR. The PHATTER survey's three bricks (54 individual fields of view) span the inner disk of M33, extending out to a galactocentric radius of $\sim$4~kpc. We use three types of image products from the survey. First, drizzled single-pointing ACS images were used to create optical images with synthetic clusters inserted to minimize computational effort (see \S\ref{sec:synclst}).  Second, brick-wide mosaic images for each of the six filters were used for aperture photometry of the clusters, which provide the best overlapping spatial coverage and artifact removal (i.e., chip gaps and cosmic rays).  Third, LGCS search images (see \S\ref{sec:interface}) were extracted from survey-wide optical mosaic images.  All images have an image scale of 0.05~arcsec pixel$^{-1}$ are astrometrically aligned to Gaia DR2 with 3~mas (7~mas) residuals for ACS/WFC and WFC3/UVIS (WFC3/IR), and are combined and distortion corrected using \texttt{AstroDrizzle} from the \texttt{DrizzlePac} package \citep{DrizzlePac12,Hack13,Avila15}.

In addition to the images, we also use PHATTER resolved star photometry catalogs presented in \citep{Williams21}.  This PSF photometry was measured simultaneously in all six filters using DOLPHOT\footnote{\url{http://americano.dolphinsim.com/dolphot/}}, an updated version of the HSTphot photometry package \citep{Dolphin00}.

We use the PHATTER photometry catalogs to quantify and map stellar density across the survey footprint.  Specifically, we define and use two quantities: $N_{\mathrm{MS}}$, the number of upper main sequence stars selected using a color magnitude cut of F475W $<$ 24 and F475W$-$F814W $<$ 1; $N_{\mathrm{RGB}}$, the number of bright RGB stars  defined using a polygon region in the optical color magnitude diagram (CMD) that mimics a NIR-based selection used in M31 by \citetalias{Johnson15_AP}, where F475W$-$F814W $> $1.5 and F814W brighter than $\sim$22.5.

\subsection{LGCS Interface} \label{sec:interface}

The Local Group Cluster Search (LGCS)\footnote{\url{https://www.clustersearch.org/}} is a citizen science project built and hosted on the Zooniverse\footnote{\url{https://www.zooniverse.org/}} platform. The project is a direct follow-on of the Andromeda Project \citepalias{Johnson15_AP}, but was built using the Zooniverse's Project Builder\footnote{\url{https://www.zooniverse.org/lab}} platform tools rather than being built as a custom project-specific website. The Project Builder platform allowed the research team to build and configure the project without the effort or assistance of the Zooniverse web development team, though LGCS has fewer custom features than the Andromeda Project (e.g., no interactive walk-through of interface tools during project tutorial). The main capabilities provided by the Zooniverse platform, however, remain the same: an interactive user interface that enables image annotation, web hosting for the project page and image data, subject image selection and queuing, feedback confirming the correct identification of synthetic clusters in the images (see \ref{sec:synclst}), and storage of classification responses.

The scope of the LGCS project extends beyond the search of PHATTER imaging presented here; the project hosts a visual cluster search of SMASH \citep{Nidever17} imaging of the Large and Small Magellanic Clouds, and will expand to additional datasets in the future. This study focuses solely on the results of the PHATTER M33 search, while results from other LGCS searches will be published separately in future work.

\begin{figure*}
    \centering
    \includegraphics[width=0.75\textwidth]{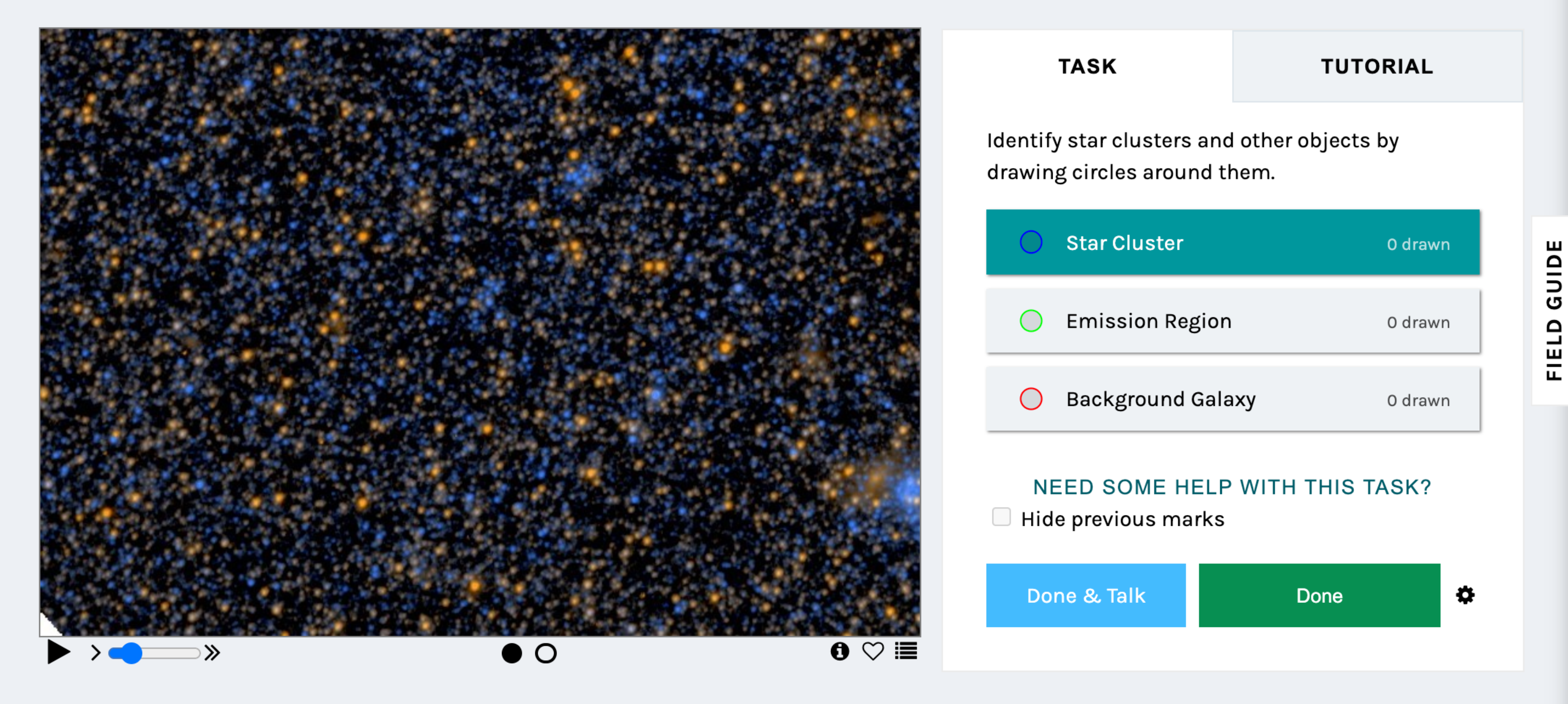}
    \caption{Screenshot of the Local Group Cluster Search annotation interface.}
    \label{fig:website}
\end{figure*}

Volunteers who participate in the LGCS search of the PHATTER data use a simple annotation interface to mark objects of interest, as shown in Figure~\ref{fig:website}. Specifically, participants are asked to mark star clusters, background galaxies, and nebulous emission regions using a circular tool with adjustable size.  Users click the center of an object in the image and drag outward to set a circular marker's radial size. Two images are shown to participants: a color image constructed from F475W and F814W bands, and an inverted grayscale F475W image.  The two-band composite image provides important color information, while the inverted single-band image provides high contrast to improve the detection of faint clusters. Individual $\sim$36$\times$25 arcsec ($\sim$150$\times$100 pc) subimages are extracted from survey-wide drizzled mosaic images. These subimages spatially overlap by 100 pixels (5 arcsec) to minimize edge effects on search results.

In addition to the image annotation interface, the LGCS project also hosts helpful resources for volunteers.  Participants are presented a tutorial for instruction about the task during their first visit to the classification page, which includes a short demonstration video and text instructions.  Volunteers can also access a number of ``About'' pages that summarize the project's research goals and background information, a field guide that provides detailed examples of target objects, and a forum called ``Talk'' to facilitate interactions between with the research team and project participants.

\subsection{Data Collection and Classification Statistics} \label{sec:datacollection}

The LGCS project collected image classifications from 8 January 2019 to 28 February 2019. During this time, LGCS volunteers submitted 269,645 classifications, where one classification denotes a participant's response specifying the location and size of any objects they identify in an image. Each search image was classified by at least 60 unique volunteers.

\begin{figure}[t]
    \centering
    \includegraphics[width=8.5cm]{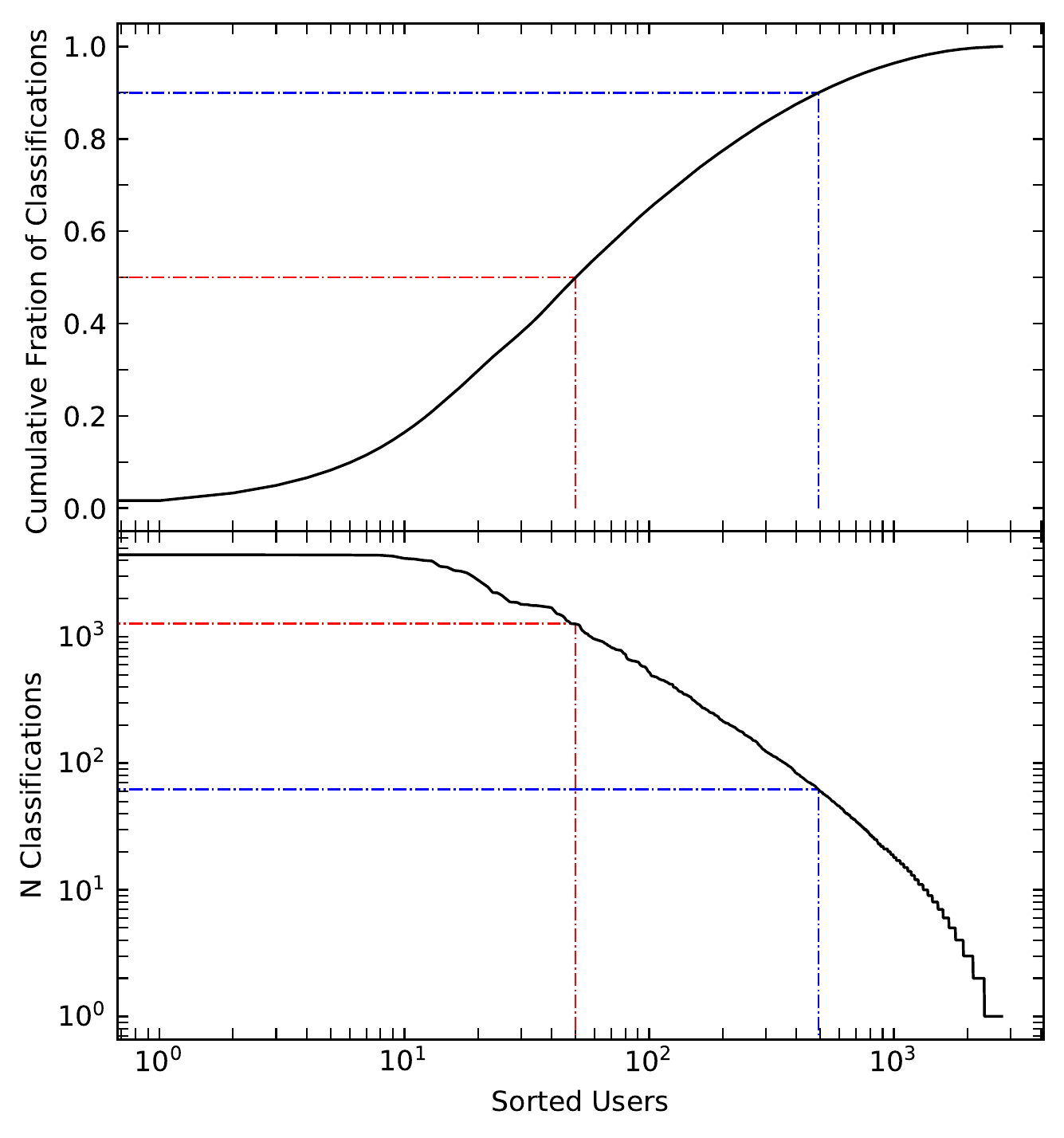}
    \caption{Classification statistics of individual LGCS users, sorted in decreasing order of total submitted classifications. Top: The cumulative fraction of the LGCS's $\sim$269 thousand classifications. Bottom: The number of classifications submitted by the Nth ranked volunteer. The red dashed-dotted lines illustrate that 50\% of all classifications were performed by the 50 most active volunteers who each contributed at least 1267 classifications. The blue dashed-dotted lines illustrate that 90\% of all classifications were performed by 463 volunteers who each contributed at least 62 classifications.}
    \label{fig:userstats}
\end{figure}

A total of 2757 users participated in the LGCS project, 1517 of whom participated as registered users using a Zooniverse account. Contributions from individuals who were not logged in are identified and grouped via an anonymized hash of the participant's IP address. We note that while non-registered users represent 45\% of volunteers by number, these users only contributed 8\% of the total classifications. Beyond this difference between registered and non-registered users, the distribution of effort across the pool of participants varies significantly, as shown in Figure~\ref{fig:userstats}. The median number of classifications per person is nine (nineteen for registered users), but half of all classification effort was provided by the 50 most active volunteers --- each of whom contributed at least 1267 classifications.  This behavior is consistent with trends seen for the Andromeda Project \citepalias{Johnson15_AP} and other Zooniverse projects \citep{Spiers19}, where a relatively small group of active volunteers contribute a bulk of the total effort.

\pagebreak 
\subsection{Synthetic Cluster Generation} \label{sec:synclst}

A key part of our analysis is incorporating synthetic clusters with known ages, masses, and radii into the visual search.  Synthetic clusters were inserted into the same LGCS images used for the cluster search, but were analyzed separately. The synthetic clusters were created following a procedure similar to the one used by \citetalias{Johnson15_AP}.  In short, a synthetic cluster's individual member stars were drawn with masses following a \citet{Kroupa01} IMF and stellar properties for a specified age using PARSEC 1.2S + COLIBRI PR16 isochrones \citep{Bressan12,Marigo17}.  Spatial positions were drawn from a \citet{King62} profile with a specified effective radius, $R_{\mathrm{eff}}$, and a concentration ($R_{\mathrm{tidal}}/R_{\mathrm{core}}$) of 30.  The sample of synthetic clusters were created with the following properties:
\begin{enumerate}
    \item Ages were drawn randomly from a grid of log(Age/yr) values ranging from 6.6 to 10.1 incremented every 0.05 dex.
    \item Masses were drawn randomly from a continuous uniform distribution of log(Mass/\Msun) values ranging from 2.0 to 5.0.
    \item A fixed Solar metallicity ($Z$=0.0152) was assumed for ages younger than 5 Gyr. For older ages, the metallicity was randomly drawn from a set of five discrete values ($Z$=[0.0152,0.005,0.0015,0.0005,0.00015]) so that the sample of older synthetic clusters would span a metallicity range resembling that of Galactic and extragalactic globular clusters.
    \item Extinctions were drawn from an exponential $A_V$ distribution which ranges from the foreground Milky Way extinction value of 0.11 to 3.0 mags, following the expression: $P(A_V) \propto e^{-(A_V / 1.34)}$; this is the same distribution used for M31 synthetic clusters by \citetalias{Johnson15_AP}.  
    \item Effective radii ($R_{\mathrm{eff}})$ were drawn from the distribution of measured values obtained for M31 clusters by \citetalias{Johnson15_AP}, but biased to larger $R_{\mathrm{eff}}$ values to ensure sufficient number statistics of diffuse clusters in the high-$R_{\mathrm{eff}}$ tail for completeness determination purposes.
\end{enumerate}

After creating a parent population of artificial clusters, we chose a subset to insert into LGCS images. This selection was based on cluster magnitude and age, and it was designed to produce a sample of synthetic clusters that spans the full range of detectability, from easily detected to undetectable. Specifically, we adopt the following magnitude limits: $18.5 < m_{\mathrm{F475W}} < 22$ for $6.6 < \log(\mathrm{Age/yr}) < 8.0$; $19.5 < m_{\mathrm{F475W}} < 22.5$ for $8.0 < \log(\mathrm{Age/yr}) < 9.0$; $20 < m_{\mathrm{F475W}} < 22.5$ for for $9.0 < \log(\mathrm{Age/yr}) < 10.0$.

We inserted the magnitude-selected sample of synthetic clusters into F475W and F814W images using DOLPHOT. One synthetic cluster was added per LGCS search image, positioned pseudo-randomly within the image, avoiding positions within 120 pixels of the edge. Because DOLPHOT places the synthetic clusters into each individual frame before drizzling, we use single-field images as opposed to multi-image mosaics for insertion to minimize computational complexity.  Insertion locations were chosen to avoid chip edges and gaps to ensure that the synthetic images were essentially identical to the original search images. 

We created two batches of synthetic clusters, each with 848 objects for a total of 1696 synthetic clusters. The first batch was randomly assigned to LGCS images spanning the entire PHATTER survey footprint, resulting in a diverse set of cluster-image pairs across the full range of galactic environments. The second batch was assigned spatial locations in a targeted manner, such that young clusters ($<$100 Myr) were placed in regions of the footprint with a high density of bright, blue stars, as defined by their high $N_{\mathrm{MS}}$ values ($N_{\mathrm{MS}} > 1200$ stars per search image).  The remaining older synthetic clusters were distributed across the remaining fields with lower $N_{\mathrm{MS}}$ values.  The targeted placement of this second batch ensures sufficient numbers of young synthetic clusters fall within young star-forming regions, safeguarding our ability to derive catalog completeness for the key population of young star clusters.

\section{Catalog Construction} \label{sec:catcon}

The process of converting LGCS image classifications into a star cluster catalog involves combining 60 independent classifications from each image into a consensus result regarding the presence, location, and size of candidate clusters (and other objects).  This target number of classifications per image is selected as a balance of keeping statistical errors on classification results low and maintaining a reasonable total runtime for the project.

The first step in this process is to compile and combine the candidate identifications made by LGCS participants.  We merge identifications following the procedure described in detail in Appendix A of \citetalias{Johnson15_AP}: we aggregate markings for each individual search image by clustering marker centers and merging overlapping candidates, then we combine the per-image lists of identifications into a survey-wide data product by running a spatial match to merge duplicate candidates in regions of overlapping image coverage.

We use the fraction of classifications where the object is detected as the principal indicator of significance.  We compute four fractional metrics to characterize each candidate: 
\begin{enumerate}
    \item $f_{\mathrm{view}}$ is the fraction of total classifications where a candidate is identified as any class of object.
    \item $f_{\mathrm{cluster}}$ is the fraction of total classifications where a candidate is identified as a star cluster
    \item $f_{\mathrm{galaxy}}$ is the fraction of total classifications where a candidate is identified as a background galaxy
    \item $f_{\mathrm{emission}}$ is the fraction of total classifications where a candidate is identified as an emission region.
\end{enumerate}
These quantities are related by:
\begin{equation}
    f_{\mathrm{view}} = f_{\mathrm{cluster}} + f_{\mathrm{galaxy}} + f_{\mathrm{emission}}
\end{equation}
We note that the definitions of these metrics differ slightly from those used by \citetalias{Johnson15_AP}, such that all four metrics are normalized by the total number of available image classifications.

The aggregation process produced a set of 10926 unique identifications. This total number includes many low significance objects, with only 4780 candidates having $f_{\mathrm{view}} \ge 0.1$. For this paper, we focus primarily on the cluster candidates; please see Appendix \ref{sec:app_othercat} for discussion of background galaxy and emission region results.  

\begin{figure}
    \centering
    \includegraphics[width=8.5cm]{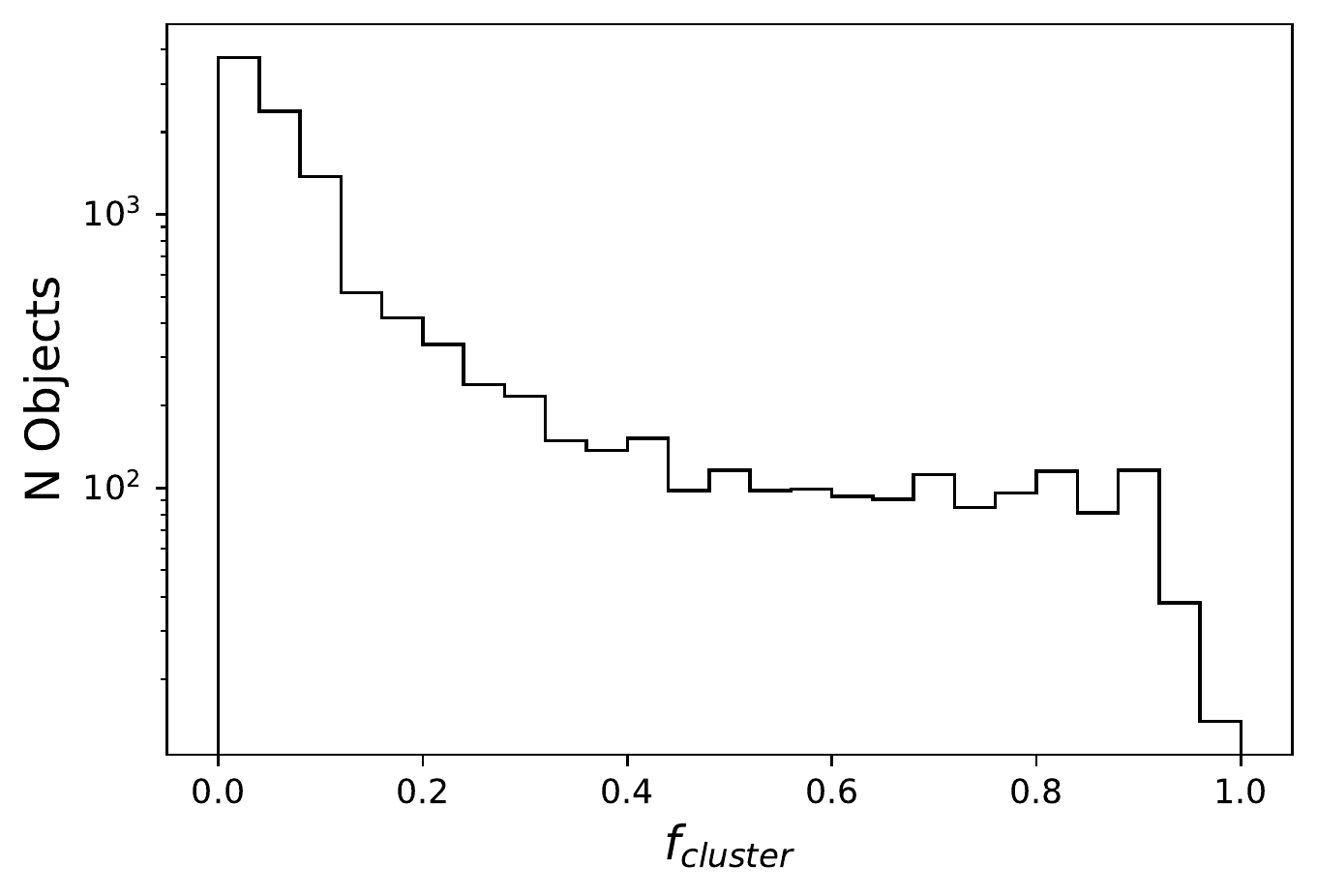}
    \caption{Distribution of $f_{\mathrm{cluster}}$ values for our full sample of cluster candidates.}
    \label{fig:f_clst_hist}
\end{figure}

We present a histogram of $f_{\mathrm{cluster}}$ values in Figure~\ref{fig:f_clst_hist}.  Our visual inspection of cluster candidates confirmed that as $f_{\mathrm{cluster}}$ decreases, the quality of the cluster candidates is lower.  We find that for $f_{\mathrm{cluster}} > 0.6$, there are very few contaminants; among 841 candidates, only one has $f_{\mathrm{galaxy}} > 0.1$, and that object is eliminated by subsequent weighted cuts (see \S\ref{sec:userweighting}).

\subsection{User Weighting}
\label{sec:userweighting}

\begin{figure*}[ht]
    \centering
    \includegraphics[width=0.75\textwidth]{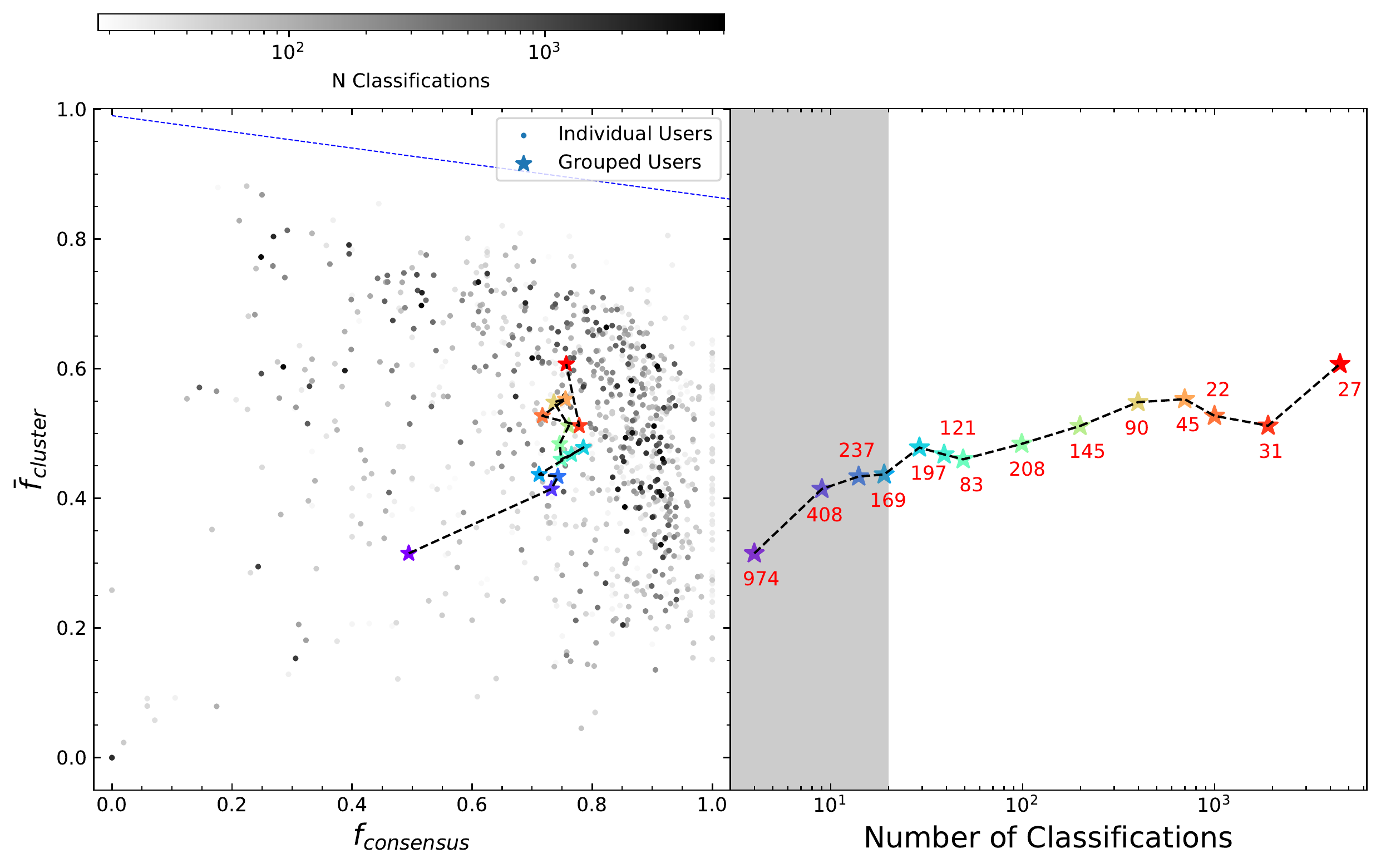}
    \caption{
    {\em Left:} $f_{\mathrm{consensus}}$ and $\overline{f}_{\mathrm{cluster}}$ user metric scores used for weighting. Circular points represent metrics for individuals with $>$20 classifications; point shading denotes volunteer classification count, where darker colors highlight individuals with more classifications. Star markers represent the average metrics for a group of users, binned in groups according to classification count. The blue dashed line represents the maximum $\overline{f}_{\mathrm{cluster}}$ a volunteer can achieve based on the fraction of high-quality consensus clusters identified.
    {\em Right:} Star markers correspond to the same binned user groups presented in the left panel, but now showing the average number of classifications for each group on the x-axis. The red number next to each star is the number of users represented in each respective bin. For volunteers with $<$20 classifications (gray shaded region) users were weighted according to aggregate user metric values calculated for the binned group rather than using individual values.
    }
    \label{fig:user_metrics}
\end{figure*}

While the $f_{\mathrm{cluster}}$ metric assumes that each volunteer is equally skilled at identifying star clusters, multiple citizen science projects \citep[e.g.,][]{Willett13, Jayasinghe19, Eisner21} have found that weighting volunteer responses according to their task performance can increase sample quality and decrease catalog contamination as a function of completeness. To maximize the usefulness of LGCS volunteer contributions, we follow the methodology of \citetalias{Johnson15_AP} and weight classifications based on the volunteer's performance identifying star clusters.  In this section, we demonstrate that employing user weighting significantly improves the resulting cluster catalog.

We calculate user weights based on the agreement between the classifications of the user and the entire set of LGCS participants. We use two separate weights: one for detections, objects the user identified; one for non-detections, objects the user did not identify. A user's detections are weighted according to the average $f_{\mathrm{cluster}}$ of all cluster identifications made by the individual, such that classifications by those who tend to identify good candidates with higher $f_{\mathrm{cluster}}$ are assigned greater weights than those who identify worse candidates with lower $f_{\mathrm{cluster}}$.  A user's cluster non-detections are weighted according to the fraction of high-quality clusters ($f_{\mathrm{cluster}} > 0.6$) that the user sees and detects, such that classifications by those who rarely miss good clusters carry greater weight than those who are more selective and identify fewer clusters.

We note that volunteer classifications of synthetic cluster images are omitted from user metric calculations.  This omission ensures that user classification metrics are based only on real data, and that catalog completeness is not biased due to user weighting.

We calculate two user metrics for all volunteers to quantify the detection and non-detection behaviors described above: $\overline{f}_{\mathrm{cluster}}$, the average $f_{\mathrm{cluster}}$ of all clusters a user identifies, and $f_{\mathrm{consensus}}$, the fraction of high-quality ($f_{\mathrm{cluster}} > 0.6$) clusters a user saw that they identified.  Figure~\ref{fig:user_metrics} shows these user metrics for all volunteers who contributed more than 20 classifications. Many users lie toward the upper right corner of the plot, representing users who excel in both user metrics.  In contrast, users in the top left are conservative classifiers who identify clusters with high $f_{\mathrm{cluster}}$, but miss a significant fraction of commonly-identified clusters.  Those in the bottom right are liberal classifiers who include all good clusters in their identifications, but at the expense of also including lower quality clusters as well.

We examine trends in user behavior by grouping users into bins according to their total classification count, with the average user metrics of each group plotted as colored stars in Figure~\ref{fig:user_metrics}. Volunteers with higher classification counts (redder points) tend to have higher user metrics scores, which may indicate that volunteers become more skilled on average as they classify an increasing number of images.

Because an individual's user metrics become noisy at small numbers of classifications, we replace the metrics of users with $\leq$20 classifications with the aggregate values of $\overline{f}_{\mathrm{cluster}}$ and $f_{\mathrm{consensus}}$ obtained for their binned groups.  The use of aggregate metrics and weights for users with low classification counts has little impact on the catalog results due to the small percentage of total classifications contributed by these individuals (see Figure~\ref{fig:userstats}).

To convert a volunteer's user metric results into a classification weight, we adopt a generalized logistic function:
\begin{equation}
    W(x) = B \times \left( A + \frac{1}{1 + e^{-m_{\mathrm{logistic}}(x-b_{\mathrm{logistic}})}} \right), 
\end{equation}
where $x$ represents $\overline{f}_{\mathrm{cluster}}$ for detection weights and $f_{\mathrm{consensus}}$ for non-detection weights, while $m_{\mathrm{logistic}}$ and $b_{\mathrm{logistic}}$ are the slope and position of maximum growth of the logistic curve. The coefficients $A$ and $B$ are normalization constants set such that $W$ varies between 0 and 1 over the $x$ interval [0, 1].  We seek to identify values of $m_{\mathrm{logistic}}$ and $b_{\mathrm{logistic}}$ for detection and non-detection weights that maximize cluster catalog completeness and minimize contamination.

To fit for an optimal weighting scheme, we first compile a set of ``expert'' ratings to use as a reference when computing completeness and contamination metrics for a given set of weighting parameters.  A group of 4 co-authors visually inspected clusters and scored them on a scale of 1--3: 1 is a definite cluster, 2 is a possible cluster, and 3 is a non-cluster.  Four co-authors ranked all marginal cluster candidates ($0.35 < f_{\mathrm{cluster}} < 0.5$) where we expect the greatest variety in quality.  Additionally, one co-author ranked a broader range of candidates ($f_{\mathrm{cluster}} \gtrsim 0.25$) to confirm that objects with $f_{\mathrm{cluster}} < 0.35$ were low quality identifications. The average of these ranks, $S_{\mathrm{expert}}$, is then used to categorize clusters and contaminants. Clusters with $S_{\mathrm{expert}} < 1.5$ were declared good clusters, and $S_{\mathrm{expert}} > 2.5$ were considered contaminants.

\begin{figure}
   \centering
    \includegraphics[width=8.5cm]{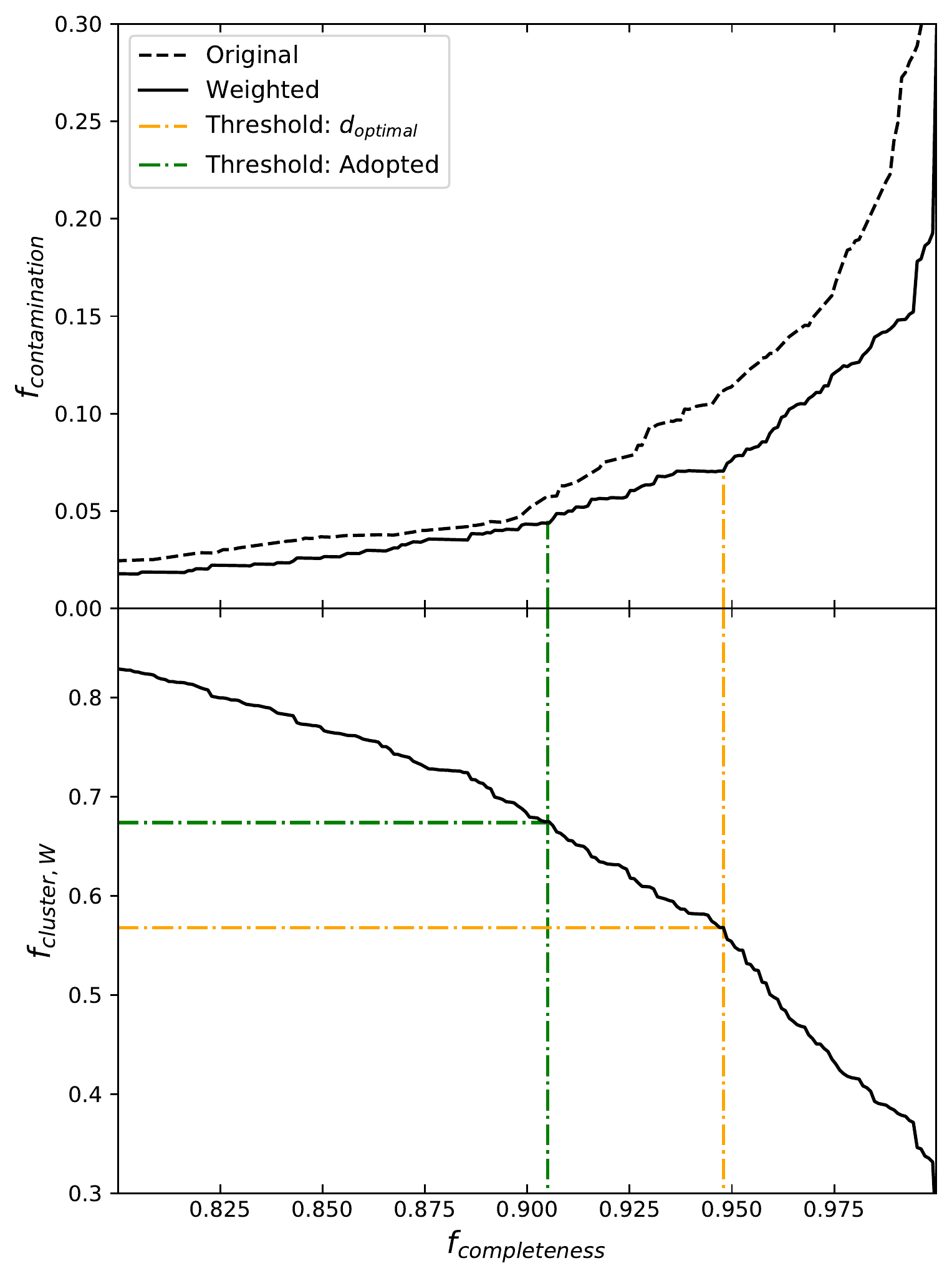}
    \caption{Top: catalog completeness versus contamination for the original, unweighted sample (dashed black line), and the optimal user weighting system (solid black line). The orange dashed line represents the $d_{\mathrm{optimal}}$ threshold, where $d_{\mathrm{optimal}}$ is the distance from this point to the bottom right corner of the plot. The green line shows the adopted threshold chosen to reduce catalog contamination. Bottom: catalog completeness versus $f_{\mathrm{cluster, W}}$, which provides reference to where each threshold is drawn.}
    \label{fig:comp_vs_cont}
\end{figure}

Using the expert ratings, we construct a completeness versus contamination curve for our unweighted sample by varying the $f_{\mathrm{cluster}}$ threshold from 0 to 1, as shown in Figure~\ref{fig:comp_vs_cont}.  We define the minimum distance from the curve to the lower left corner of this plot (i.e., an optimal sample with 100\% completeness and no contaminants) as $d_{\mathrm{optimal}}$, and use this metric to evaluate, rank, and optimize the adjustable weighting parameters.

We conduct an iterative, grid-based search for an optimal set of detection and non-detection weighting function parameters that minimize the $d_{\mathrm{optimal}}$ metric for the completeness versus contamination curve.  For each grid point, we calculate detection and non-detection user weights from $\overline{f}_{\mathrm{cluster}}$ and $f_{\mathrm{consensus}}$ user metrics, respectively, using logistic functions with specified values of $m_{\mathrm{logistic}}$ and $b_{\mathrm{logistic}}$ parameter values to make the metric-to-weight transformation.  We then compute user-weighted $f_{\mathrm{cluster}}$ values, $f_{\mathrm{cluster, W}}$, for each cluster candidate, construct a completeness vs.\ contamination curve, and calculate an associated $d_{\mathrm{optimal}}$.

We identify the set of logistic function parameters that minimize $d_{\mathrm{optimal}}$, and thus produce an optimal cluster catalog that maximizes completeness and minimizes contamination. We find that the following weighting parameters produce the best weighted catalog: detection weight parameters of $(m_{\mathrm{logistic}}, b_{\mathrm{logistic}}) = (30.0, 0.45)$; non-detection weight parameters of $(m_{\mathrm{logistic}}, b_{\mathrm{logistic}}) = (30.0, 1.1)$. The optimally weighted catalog has significantly lower contamination as a function of completeness than the unweighted catalog, as shown in Figure~\ref{fig:comp_vs_cont}, demonstrating that the application of user weights improved the quality of the cluster catalog we produced.

\subsubsection{Catalog Threshold Selection}
\label{sec:threshold}

With the user weighting parameters fixed, we move on to selecting a catalog threshold. \citetalias{Johnson15_AP} choose a $f_{\mathrm{cluster, W}}$ cutoff that corresponds to the point on the completeness versus contamination curve where $d_{\mathrm{optimal}}$ is minimized.  The minimum $d_{\mathrm{optimal}}$ point corresponds to a $f_{\mathrm{cluster, W}}$ threshold of 0.568, 94.8\% completeness, and 7.1\% contamination, indicated by the orange lines in Figure~\ref{fig:comp_vs_cont}.

We note two key differences between the \citetalias{Johnson15_AP} and LGCS completeness versus contamination curves: $f_{\mathrm{contamination}}$ values for LGCS are smaller by approximately a factor of 2; the original unweighted curve (and the weighted curve to a lesser degree) shows a distinct change in slope behavior at $f_{\mathrm{completeness}} \sim 0.9$ in Figure~\ref{fig:comp_vs_cont}.  In addition, a qualitative evaluation of the $d_{\mathrm{optimal}}$-based threshold concluded that the resulting cluster sample includes a higher number of contaminants than desired, leading us to reevaluate our choice of $f_{\mathrm{cluster, W}}$ threshold.

Based on the poor assessment of the initial $f_{\mathrm{cluster, W}}$ threshold, we seek an alternative, more conservative catalog limit.  We target a greater $f_{\mathrm{cluster, W}}$ value that corresponds to a point on the weighted completeness versus contamination curve near the transition in slope at $f_{\mathrm{completeness}} \sim 0.9$.  We find that by applying a factor of 2 scaling to the $f_{\mathrm{contamination}}$ component of the $d_{\mathrm{optimal}}$ distance calculation, motivated by the $\sim$2x scaling difference between the LGCS and \citetalias{Johnson15_AP} curves, we identify a viable threshold that meets all the above criteria. The resulting $f_{\mathrm{cluster, W}}$ catalog threshold is 0.674, which corresponds to 90.5\% completeness and 4.4\% contamination, indicated by the green lines in Figure~\ref{fig:comp_vs_cont}.

\begin{figure}
    \centering
    \includegraphics[width=8.5cm]{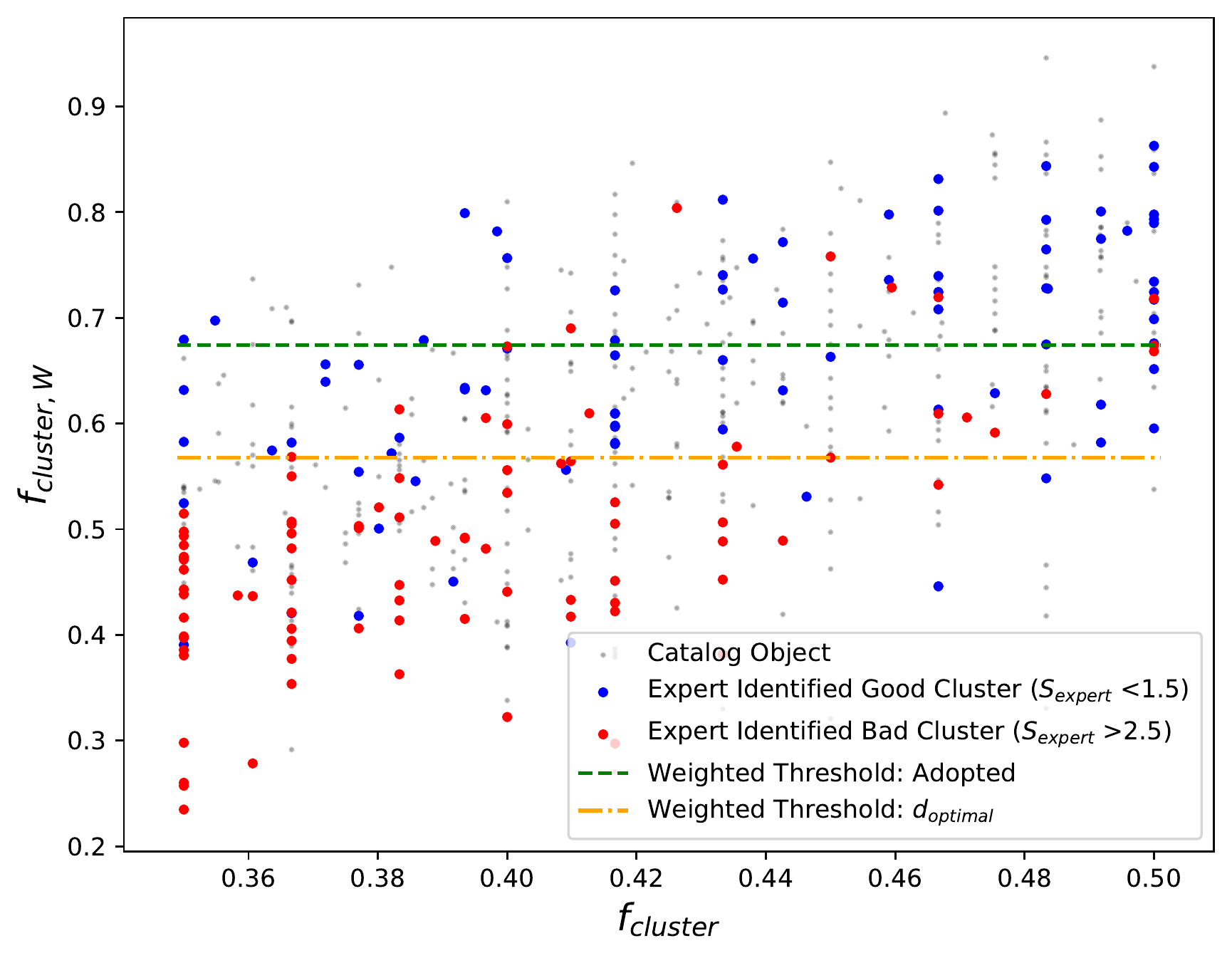}
    \caption{Weighted and unweighted $f_{\mathrm{cluster}}$ values are plotted for the subset of expert ranked clusters. The blue dots are expert identified good clusters, and the red dots are expert identified contaminants. The dashed green line represents the $f_{\mathrm{cluster, W}}$ catalog threshold of 0.674, while the orange dashed line represents the $f_{\mathrm{cluster, W}}$ catalog threshold of 0.568. We prefer the higher adopted $f_{\mathrm{cluster, W}}$ threshold as it rejects a larger number of expert identified bad cluster candidates, as seen by the larger number of red points that that fall below the green horizontal line.}
    \label{fig:weights_work}
\end{figure}

We demonstrate the impacts of our weighting system in Figure~\ref{fig:weights_work} for the expert-classified subsample of cluster candidates with $0.35 \leq f_{\mathrm{cluster}} \leq 0.5$ where weighting and threshold selection has the greatest impact. Expert identified good clusters ($S_{\mathrm{expert}} < 1.5$) are plotted in blue, and expert identified bad clusters ($S_{\mathrm{expert}}$ > 2.5) are shown in red. This plot shows how the weighted $f_{\mathrm{cluster}}$ system is more effective at separating good candidates from bad candidates than the unweighted system, due to the improved separation of blue and red points by horizontal lines of constant $f_{\mathrm{cluster, W}}$ over vertical lines of constant $f_{\mathrm{cluster}}$. We can also see that the number of bad candidates that are rejected by the higher, adopted $f_{\mathrm{cluster, W}}$ threshold is larger than the number rejected by the original, unscaled $d_{\mathrm{optimal}}$-based $f_{\mathrm{cluster, W}}$ threshold, justifying our choice of the more conservative catalog threshold.


\begin{figure}
    \centering
    \includegraphics[width=0.44\textwidth]{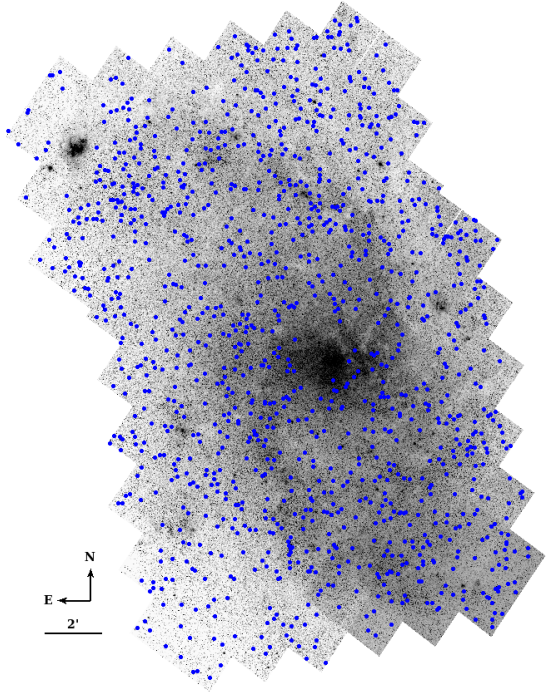}
    \caption{Spatial distribution of PHATTER star clusters (blue) overlaid on a F475W survey-wide mosaic image.}
    \label{fig:map}
\end{figure}

\begin{figure}
    \centering
    \includegraphics[width=0.34\textwidth]{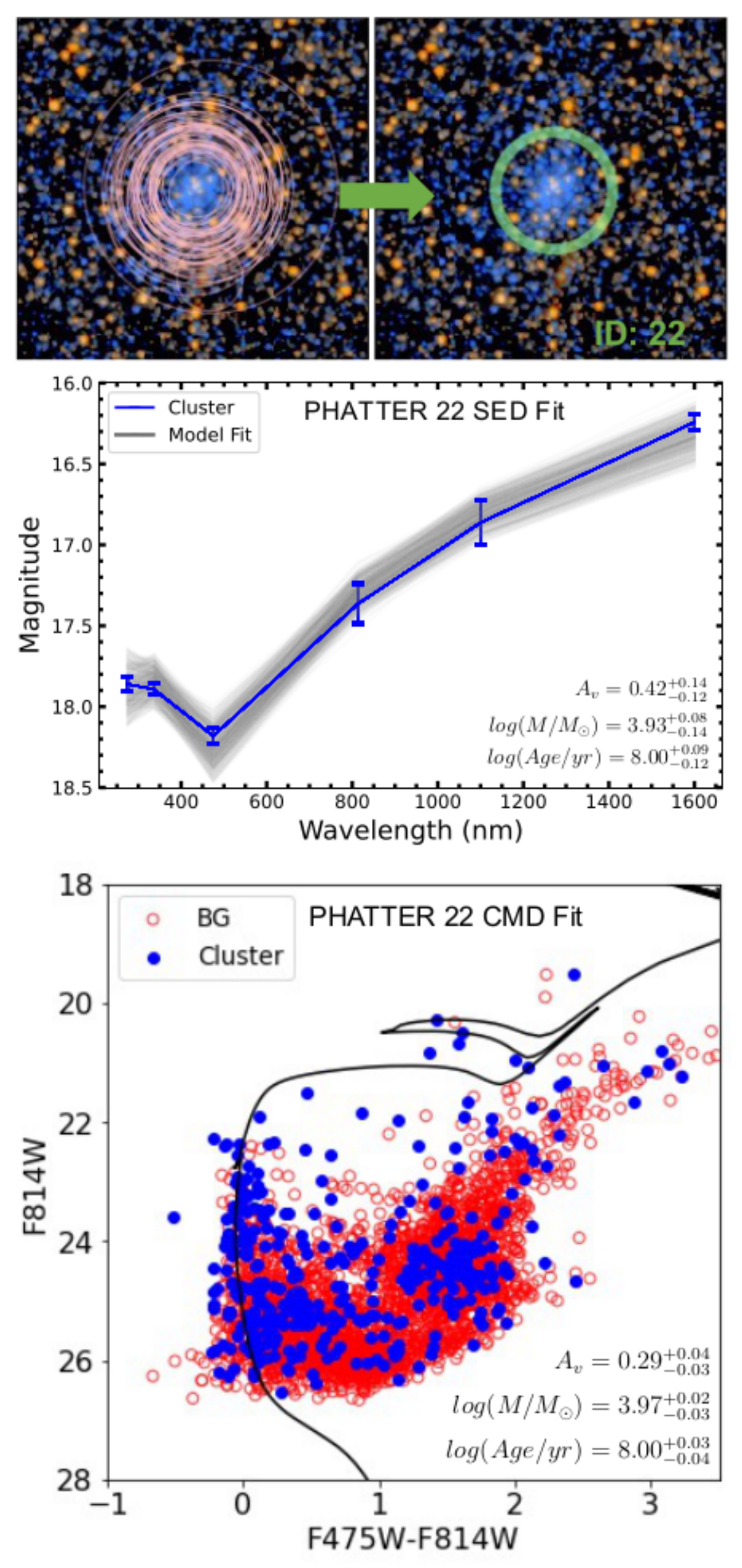}
    \caption{An example cluster: PHATTER 22. Top: Two $12\times12$ arcsec F475W+F814W color images, where the pink circles represent user cluster identifications, and the green circle represents the final cluster aperture ($R_{\mathrm{ap}}$ of 2.04 arcsec) derived from the median radius of the individual user apertures. Middle: The PHATTER 22 SED (blue) created from six-band integrated photometry. The results of SLUG SED fitting for PHATTER 22 are printed in the lower right (50th percentile value with 16th-84th percentile confidence interval), and gray lines show the 100 best-fit SLUG models. Bottom: The PHATTER 22 cluster CMD, where stellar photometry for sources within the cluster aperture (blue) are accompanied by surrounding photometry of field stars (red). CMD fitting results from \citet{Wainer22} are listed in the lower right, and a stellar isochrone (black) depicting the cluster's best fit properties is overplotted.}
    \label{fig:examplecluster}
\end{figure}

\section{PHATTER Star Cluster Catalog} \label{sec:catalog}

We apply the catalog construction techniques and user weighting described in Section \ref{sec:catcon} to derive a final cluster catalog. We present a sample of 1214 clusters that were selected using an $f_{\mathrm{cluster, W}}$ threshold of 0.674, which reflects a conservative selection of clusters that minimizes contamination by non-cluster candidates. Table \ref{tab:cat} reports cluster position, radius, and classification metrics (including weighted $f_{\mathrm{cluster, W}}$) for each of our cataloged clusters. We show the spatial distribution of the clusters in Figure~\ref{fig:map}.

We also present information for an additional 3566 candidate identifications with $f_{\mathrm{cluster, W}} < 0.674$ and $f_{\mathrm{view}} \ge 0.1$ in Appendix \ref{sec:app_othercat}, allowing catalog users to make alternative choices of catalog $f_{\mathrm{cluster, W}}$ thresholds based on the specific needs and requirements of a particular science use case.

We present an example cluster, PHATTER 22, in Figure~\ref{fig:examplecluster} to illustrate the data available for the PHATTER cluster sample.  Optical cutout images display individual cluster identifications made by LGCS volunteers as well as the final cluster aperture for the cluster, and the images show that star clusters appear as collections of individually-resolved members stars in PHATTER imagery.  The survey's panchromatic images produce six-band SEDs from integrated light photometry (see \S\ref{sec:photometry}) and cluster CMDs from resolved star photometry catalogs, both of which are used to fit for cluster properties (see \S\ref{sec:slugfitting} and \citealt{Wainer22}).

\movetabledown=2.3in
\begin{rotatetable*}
\begin{deluxetable*}{cccccccccccccccccccccccc}
\tabletypesize{\footnotesize}
\setlength{\tabcolsep}{0.05in}
\tablewidth{0pt}
\tablecaption{PHATTER Cluster Catalog \label{tab:cat}}

\tablehead{
\colhead{ID} & \colhead{RA (J2000)} & \colhead{DEC (J2000)} & \colhead{$R_{\mathrm{ap}}$ ($\arcsec$)} & \colhead{$R_{\mathrm{eff}}$ ($\arcsec$)} & \colhead{$f_{\mathrm{view}}$} & \colhead{$f_{\mathrm{cluster}}$} & \colhead{$f_{\mathrm{galaxy}}$} & \colhead{$f_{\mathrm{emission}}$} & \colhead{$f_{\mathrm{cluster, W}}$} & \colhead{Flags} & \colhead{$m_{\mathrm{apcor}}$} & \colhead{$m_{275}$} & \colhead{$\sigma_{275}$} & \colhead{$m_{336}$} & \colhead{$\sigma_{336}$} & \colhead{$m_{475}$ }& \colhead{$\sigma_{475}$} & \colhead{$m_{814}$} & \colhead{$\sigma_{814}$} & \colhead{$m_{110}$} & \colhead{$\sigma_{110}$} & \colhead{$m_{160}$} & \colhead{$\sigma_{160}$}
}

\startdata
1 & 23.553754 & 30.479462 &  1.90 &  0.23 & 1.0000 & 0.9667 & 0.0000 & 0.0333 & 1.0000 & \nodata & -0.00 & \nodata & \nodata & \nodata & \nodata &  18.39 &   0.02 &  17.70 &   0.04 & \nodata & \nodata & \nodata & \nodata \\
2 & 23.611876 & 30.696172 &  1.66 &  0.48 & 1.0000 & 1.0000 & 0.0000 & 0.0000 & 1.0000 & \nodata & -0.11 & $>$21.01 & \nodata &  $>$21.10 & \nodata &  19.63 &   0.06 &  18.73 &   0.08 & \nodata & \nodata & \nodata & \nodata \\
3 & 23.606064 & 30.699199 &  1.54 &  0.36 & 1.0000 & 1.0000 & 0.0000 & 0.0000 & 1.0000 & \nodata & -0.05 & 20.09 &   0.18 &  20.42 &   0.01 &  20.21 &   0.10 &  19.41 &   0.25 & \nodata & \nodata & \nodata & \nodata \\
4 & 23.434333 & 30.514082 &  1.90 &  0.49 & 1.0000 & 0.9833 & 0.0000 & 0.0167 & 0.9999 & \nodata & -0.07 & $>$21.12 & \nodata & $>$22.39 & \nodata &  19.84 &   0.11 &  18.25 &   0.05 &  17.41 &   0.05 &  16.61 &   0.05 \\
5 & 23.583846 & 30.659212 &  1.86 &  0.37 & 0.9836 & 0.9016 & 0.0000 & 0.0820 & 0.9772 & \nodata & -0.02 & 19.29 &   0.10 &  19.00 &   0.06 &  18.76 &   0.02 &  18.00 &   0.06 &  17.64 &   0.12 &  17.21 &   0.16
\enddata
\tablecomments{Table \ref{tab:cat} is published in its entirety in the electronic edition of the {\it Astrophysical Journal}.  A portion is shown here for guidance regarding its form and content.  Note that the $R_{\mathrm{ap}}$ parameter gives the median of the user-clicked radii, which we use as the aperture for the photometry measurements (see Section~\ref{sec:photometry}).}
\end{deluxetable*}
\end{rotatetable*}

\subsection{Cluster Photometry} \label{sec:photometry}

Aperture photometry for candidate clusters and other objects are measured using the same techniques and code employed by \citetalias{Johnson15_AP}, which we summarize here.  Photometric apertures for an object are centered at the mean position and extend to the median radius ($R_{\mathrm{ap}}$) drawn by LGCS participants. The sky background is measured in 10 annuli, each with an area equal to that of the photometric aperture, extending from 1.2--3.4 $R_{\mathrm{ap}}$. These sky annuli facilitate robust determinations of sky flux levels and associated uncertainties, which are important given that background variations often dominate the photometry error budget. Photometry is reported in the VEGAMAG system for the native HST bandpasses and calibrated using zeropoints obtained from relevant instrument websites for ACS\footnote{\url{https://acszeropoints.stsci.edu/}} \citep{Bohlin16}, WFC3/UVIS\footnote{\url{https://www.stsci.edu/hst/instrumentation/wfc3/data-analysis/photometric-calibration/uvis-photometric-calibration}} \citep[2017 values;][]{Deustua17} and WFC3/IR\footnote{\url{https://www.stsci.edu/hst/instrumentation/wfc3/data-analysis/photometric-calibration/ir-photometric-calibration}} (2012 values). We note that these adopted zeropoints are consistent to within 1\% (2\%) to alternative ``2020 values'' for ACS and WFC3/UVIS (WFC3/IR).

Six-band integrated photometry for the final cluster sample is presented in Table~\ref{tab:cat}, and equivalent photometry for ancillary cluster candidates is presented in Appendix \ref{sec:app_othercat}.  Measured magnitudes are reported for detections with S/N $\ge$ 3 and 3$\sigma$ upper limits are reported for non-detections. Blank entries denote cases of incomplete image coverage in that photometric passband.

The radial light profile of each cluster is measured from the F475W image, and the half-light radius, $R_{\mathrm{eff}}$, is derived through interpolation of the radial profile.  Aperture corrections are computed assuming a \citet{King62} light profile with fixed concentration ($c = R_{\mathrm{tidal}}/R_{\mathrm{core}} = 7$), scaled to match the measured $R_{\mathrm{eff}}$. When applied to the aperture magnitudes\footnote{$m_{\mathrm{total}} = m_{\mathrm{ap}}+m_{\mathrm{apcor}}$}, these corrections yield an estimate of total cluster light that accounts for flux that falls outside the photometric aperture, $R_{\mathrm{ap}}$. The median correction is $-0.04$ mag and the 25th-to-75th interquartile range spans from $-0.09$ to $-0.01$. We report the F475W $R_{\mathrm{eff}}$ measurements and aperture corrections in Table~\ref{tab:cat}.

We flag objects with large $R_{\mathrm{eff}}$ ($\ge$0.8 arcsec or $\sim$3pc) that are also bright ($m_{\mathrm{F475W}} < 19.0$) and blue ($m_{\mathrm{F336W}}-m_{\mathrm{F475W}} < -0.5$) as possible associations. Eight such objects are identified in Table~\ref{tab:cat}.

\begin{figure}
    \centering
    \includegraphics[width=0.35\textwidth]{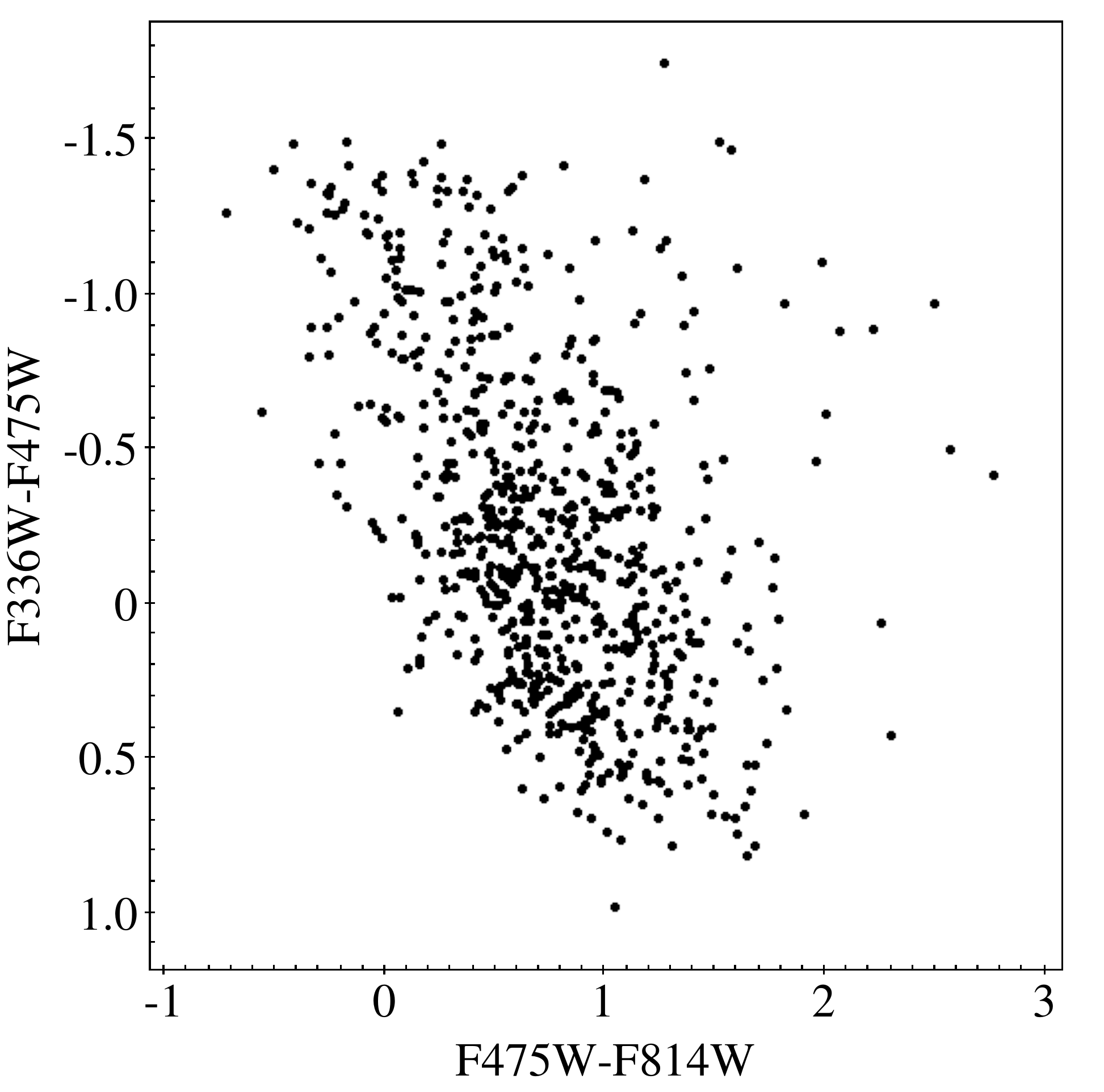}
    \includegraphics[width=0.35\textwidth]{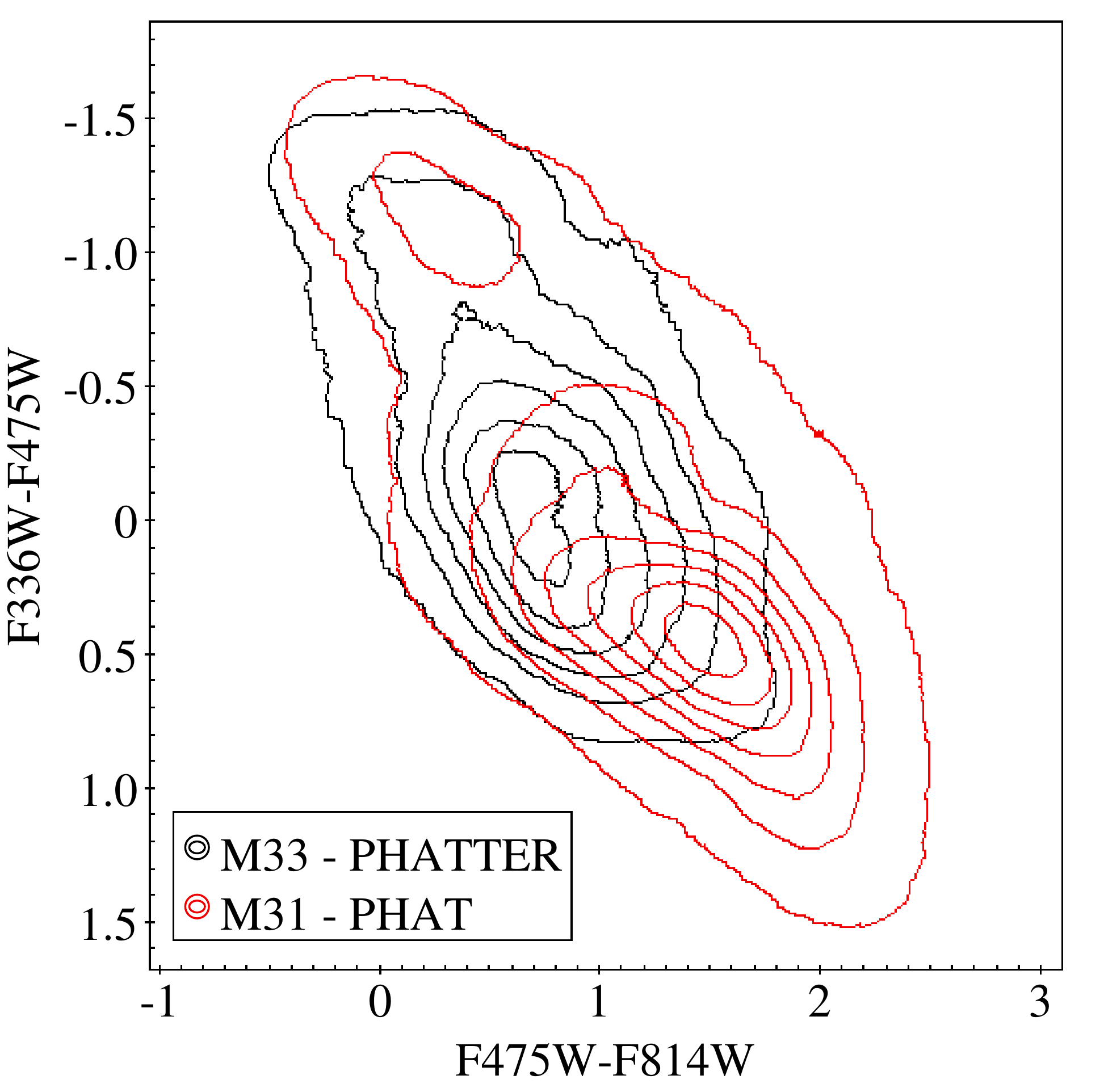}
    \caption{Top: Color-color diagram of 729 PHATTER clusters with three-band photometric detections. Bottom: Comparison of color-color distribution for PHATTER clusters in M33 and PHAT clusters in M31 \citepalias{Johnson15_AP} that shows the PHATTER clusters are significantly bluer, denoting younger cluster age.}
    \label{fig:colorcolor}
\end{figure}

We plot a UV-optical color-color diagram for the PHATTER cluster sample in Figure~\ref{fig:colorcolor}.  As expected, many of the clusters are quite blue (F336W$-$F475W $<$ 0.5), indicating young ages.  When compared to the color-color distribution of PHAT clusters from M31 \citepalias{Johnson15_AP}, we see the difference between the younger and bluer cluster population from M33 that hosts on-going star formation versus the quiescent cluster population from M31 that hosts relatively larger numbers of globular clusters and other older clusters ($>$1--3 Gyr).

\section{Catalog Completeness} \label{sec:catcompleteness}

We determine the completeness of our cluster sample by analyzing synthetic clusters inserted into LGCS images.  We analyze the classification metrics and catalog inclusion status of the synthetic cluster sample as a function of cluster luminosity, age, and mass.  We also examine the impact of cluster size and environment on cluster detection and catalog completeness.

\begin{figure*}
    \centering
\includegraphics[width=\textwidth]{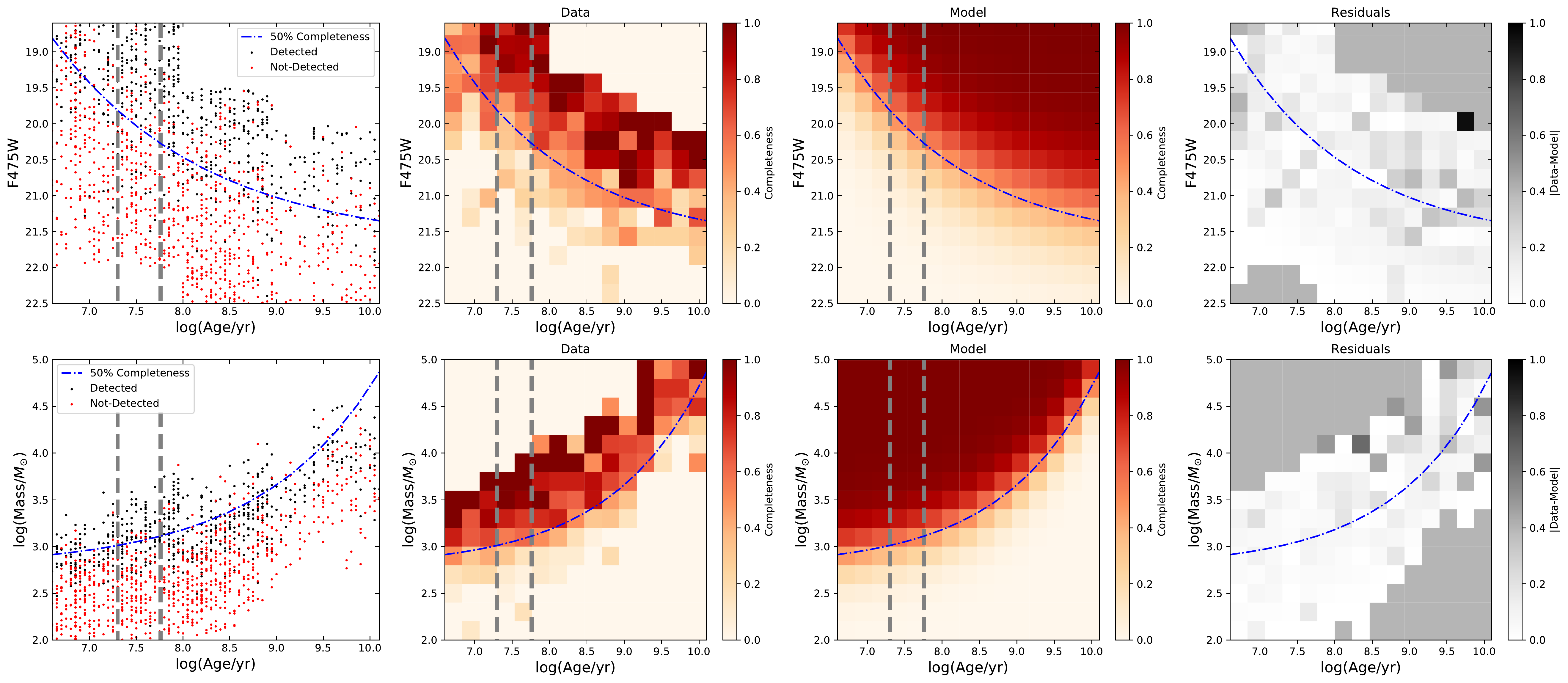}
    \caption{Completeness results based on analysis of synthetic clusters. The top panels present results as a function of F475W magnitude, the bottom panels as a function of mass. The leftmost panels present results for individual synthetic clusters: detections in black, non-detections in red. The left-middle panels show a binned version of the same results, where shading denotes the fraction of clusters detected in each bin. The middle-right panels show the completeness model fit, and the rightmost panels showing the data-model residuals. The vertical gray dashed lines in each panel highlight the age bins whose data are presented in Figure~\ref{fig:Logistic_plot}. The blue dot-dashed line in each panel represents the fitted exponential 50\% completeness model fit to the full synthetic data.}
    \label{fig:completeness}
\end{figure*}

We process classification results for the synthetic clusters through the same catalog creation pipeline used for the real LGCS cluster candidates, applying the same classification weighting and $f_{\mathrm{cluster,W}}$ detection threshold.  We present the synthetic cluster sample and detection results as an ancillary table in Appendix \ref{sec:app_othercat}. The synthetic cluster results are shown in the left panels of Figure~\ref{fig:completeness}, where black points indicate detected clusters and red points indicate non-detections. The middle-left panel shows a 2D binned representation of the results.  At higher luminosities and masses, a larger fraction of clusters are detected, and at the highest masses, nearly all synthetic clusters are detected.

The behavior of completeness with age is somewhat more complicated. Clusters at a fixed mass are more frequently detected at young ages due to their brighter luminosities. At fixed luminosity, clusters are more frequently detected at older ages due to their broader distribution of light across member stars, in contrast to young clusters that typically have a small number of very bright massive stars.

To characterize these observed trends in completeness on the age-magnitude and age-mass planes, we derive analytic formulae that can be easily applied in future modeling work. The goal here is to characterize the average completeness properties for the entire catalog, reasoning that detailed studies of specific cluster subpopulations may require a more complex completeness model than the one we present here.  

In addition to age, luminosity, and mass, dust extinction and effective radius (or, surface brightness) may also play a role. We find no need to truncate the cluster effective radius distribution, but choose to omit synthetic clusters in the high $A_V$ tail ($>$1.5 mag) when fitting for average model behavior. 

We note that the cluster environment also has an impact on completeness \citepalias[][]{Johnson15_AP}. However, because the spatial distribution of half the synthetic clusters was designed to replicate the correlation between young clusters and high $N_{\mathrm{MS}}$ regions, we expect our overall distribution of synthetic cluster environments to be comparable to the true cluster sample. We revisit the impact of cluster environment later in this section. 

\begin{figure}
    \centering
    \includegraphics[width=0.4\textwidth]{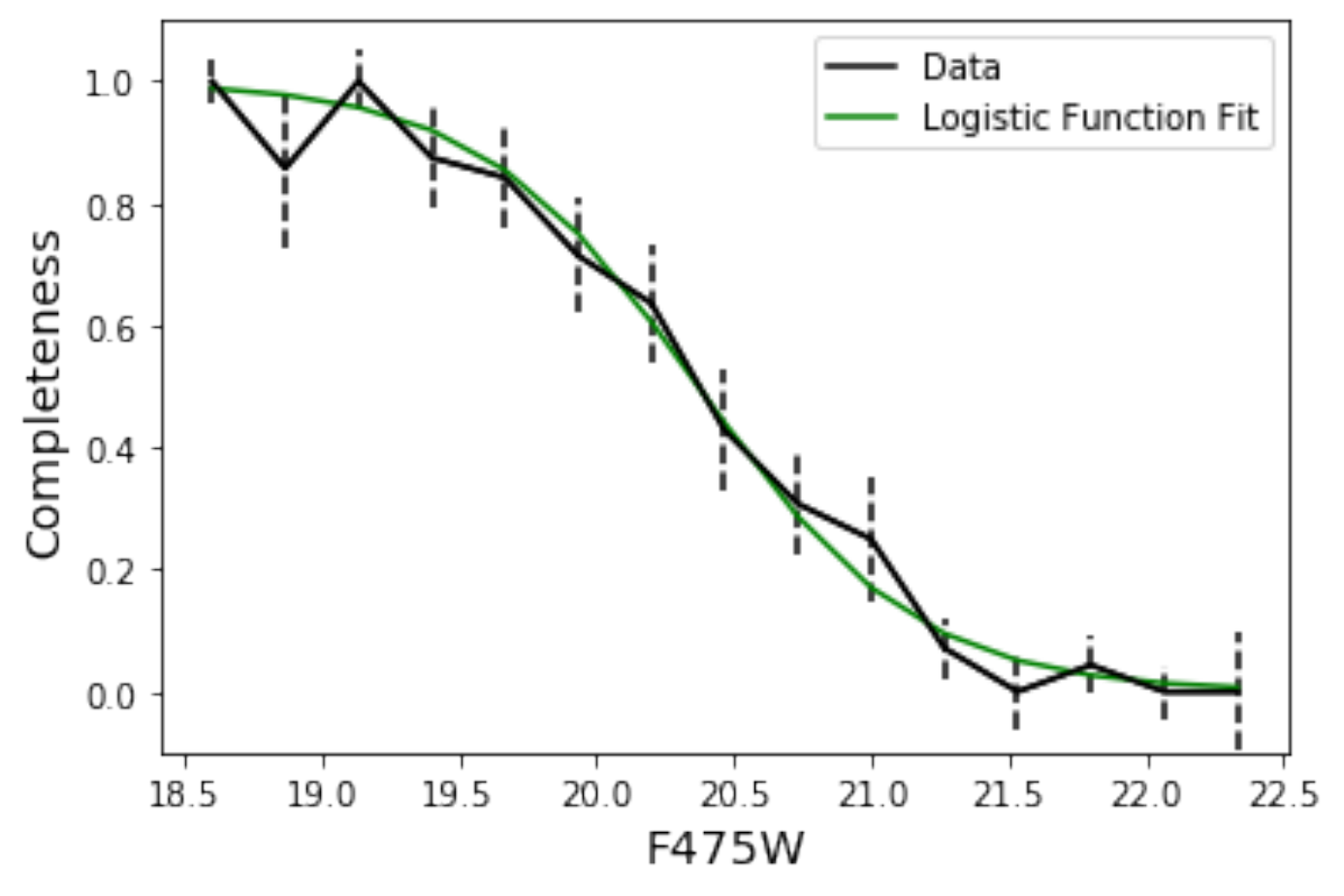}
    \includegraphics[width=0.4\textwidth]{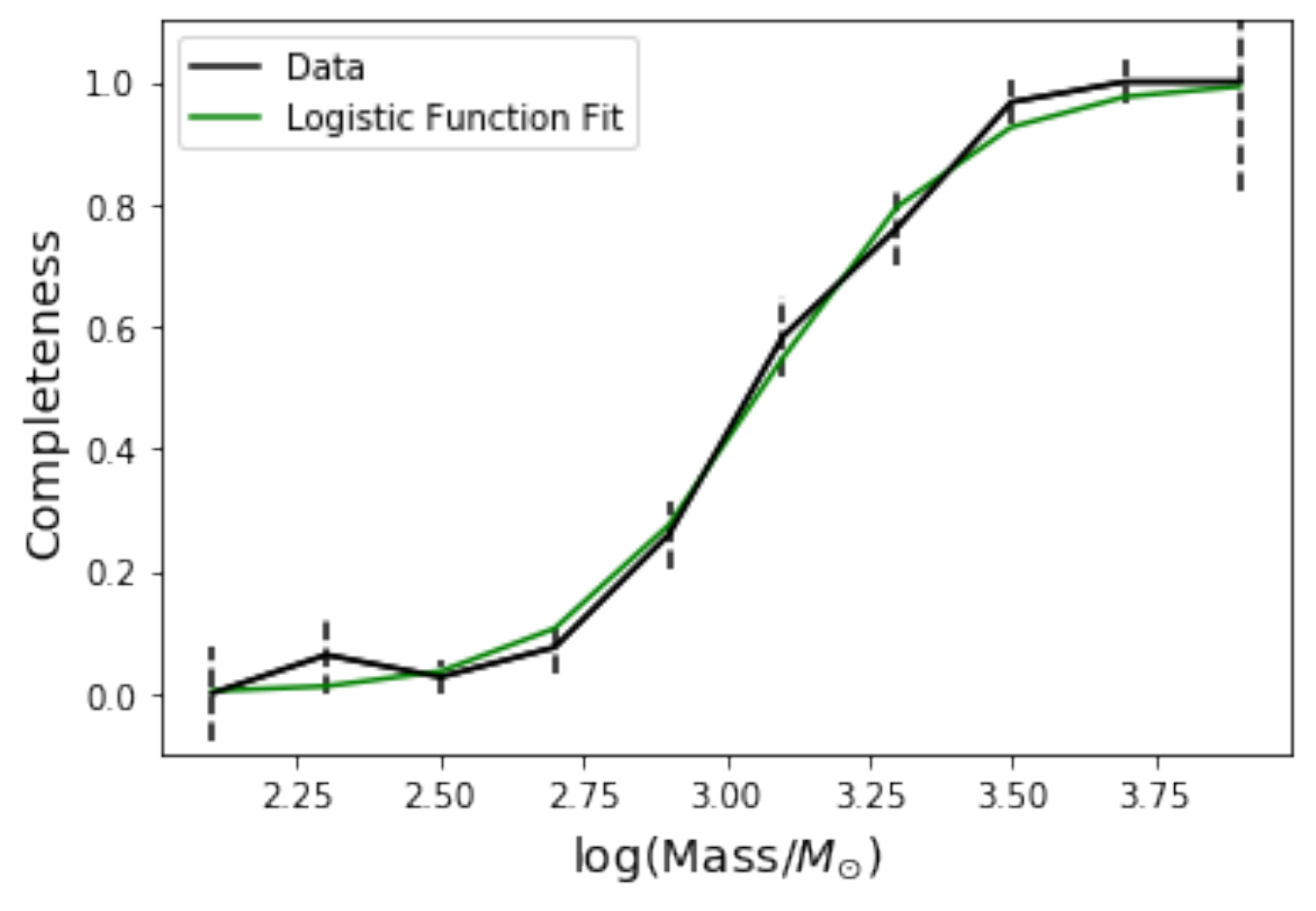}
    \caption{Logistic function fits to catalog completeness as a function of F475W magnitude (top) and mass (bottom) for synthetic clusters with ages $7.3 < {\rm log(Age/yr)} < 7.77$. Dashed lines represent the Agresti-Coull binomial proportion confidence interval for each bin.}
    \label{fig:Logistic_plot}
\end{figure}

We begin our modeling by characterizing completeness as a function of mass and F475W magnitude for bins of log(Age/yr). We use a bin width of 0.23 dex and a sliding two-bin window to improve number statistics. We use a logistic function to analytically model the completeness curve, as shown in Figure~\ref{fig:Logistic_plot}, where the black lines show the completeness data and the green line shows the best-fit logistic function. The functional form of the logistic function, $C$, is given by:
\begin{equation}
    C = ( 1 + e^{ -k ( x - M_{50}) } )^{-1}
\end{equation} 
where $k$ is the slope of the logistic function, $M_{50}$ is the 50\% completeness limit, and $x$ is the physical input parameter --- either the F475W magnitude or log($M/M_\odot$).

Based on initial fits, we find that the slope of the logistic function, $k$, is quite similar at all ages. Therefore, we fix $k$ to the median fitted value across all age bins ($-2.52$ for F475W magnitude, $5.84$ for mass) and re-fit.  

We find the age dependence of the 50\% completeness parameter, $M_{50}$, is well fit by an exponential function over the full synthetic cluster age range ($6.6 < {\rm log(Age/yr)} < 10.1$) for both mass and F475W magnitude, following the form:
\begin{equation}
    M_{50}(\tau) = a \times e^{b (\tau -\tau_{\mathrm{min}})} + c,
\end{equation}
where $\tau \equiv {\rm log(Age/yr)}$, and $\tau_{\mathrm{min}}$ is the median $\tau$ for the youngest bin, which is 6.71 for the full LGCS synthetic cluster sample and binning used here.  We note that in this formulation, the 50\% completeness in mass at $\tau_{\mathrm{min}}$ is given by $a+c$.  We fit for constants $a$, $b$, and $c$ by minimizing $\chi^2$ for the full 2D binned histogram shown in the middle-left panels of Figure~\ref{fig:completeness}.  We find the best fit exponential parameter values for mass are $(a, b, c) = (0.1455, 0.7870, 2.7810)$. For F475W magnitude the best fit values are $(a, b, c) = (-2.6767, -0.6154, 21.6834)$. The reduced $\chi^2$ for the exponential fit to the mass and F475W magnitude planes is 1.06 and 1.18, respectively. The fitted 50\% completeness exponential is presented as a dashed blue line in Figure~\ref{fig:completeness}, while the full completeness model (assuming fixed $k$ values) and its residuals relative to the synthetic cluster results are shown in the middle-right and rightmost panels of Figure~\ref{fig:completeness}.

The mass completeness behavior in the fiducial age range of $7.0 < {\rm log(Age/yr)} < 8.5$ has useful application in follow-up science analyses of young clusters, so we derive an alternate completeness model specific to this fiducial age range.  Adopting a fixed median $k$ of 6.02 and $\tau_{\mathrm{min}}$ of 7.09, we find the best fit $M_{50}$ exponential parameters are $(a, b, c) = (0.0303, 1.9899, 2.9770)$, with a reduced $\chi^2$ of 0.82. We recommend this alternative completeness model for any applications that exclude old clusters.  

We note that careful inspection of the binned F475W magnitude data in Figure~\ref{fig:completeness} (top middle-left panel) and its model residuals (top right panel) show that the youngest age bins do not neatly follow the large scale trends of the age-magnitude plane.  The fact that the age-mass plane is more well behaved at these same ages suggests this difference is likely due to stochasticity impacting the F475W magnitudes due to small number statistics of the brightest cluster members \citep[e.g.,][]{Fouesneau10,Beerman12}.  As a result, we recommend use of the mass-based completeness model whenever feasible, as we find it more reliable than our luminosity model in describing these youngest clusters.

\subsection{Additional Completeness Dependencies}

In addition to age, mass, and F475W magnitude, the spatial distribution of the cluster's member stars is another physical property that effects completeness.  Here, we use the synthetic cluster's effective radius, $R_{\mathrm{eff}}$, to parameterize cluster size and central density. For synthetic clusters with $R_{\mathrm{eff}} > 3$~pc (33\% of sample), we find a 50\% completeness limit mass that is 0.2 dex higher than for smaller clusters.  This shows that larger, diffuse clusters are systematically more challenging to detect due to the reduced central density  of cluster members, which tends to reduce the cluster's contrast against the field background.  As discussed in Section~\ref{sec:synclst}, we sampled the synthetic cluster radii from the observed cluster radii of the PHAT M31 cluster sample, but biased the distribution toward larger objects.  If we remove synthetic clusters with $R_{\mathrm{eff}} > 3$~pc, the mass completeness presented here improves by a median of 0.05 dex.  We note that for our final cluster sample, only 14\% of clusters have $R_{\mathrm{eff}} > 3$~pc.  

Additionally, local cluster environment plays a role in detection and catalog completeness. At a basic level, when the contrast between cluster and field is reduced due to an increasing density of nearby field stars of similar color, cluster detection becomes more difficult. For PHATTER images, we can characterize the local stellar density according to the number of main sequence stars per LGCS search image, $N_{\mathrm{MS}}$, calculated from counts of PHATTER photometric sources selected according to their optical CMD location.  $N_{\mathrm{MS}}$ ranges between $2.0 < log(N_{\mathrm{MS}}) < 3.75$ across the PHATTER footprint.  As $N_{\mathrm{MS}}$ increases, we see a systematic 0.7 dex increase in the 50\% mass completeness limit for the young blue star clusters, whose members are also predominately main sequence stars. As such, we recommend that future population analyses take care to account for spatial variation in catalog completeness due to stellar density, especially for trends with respect to galactocentric radius. In particular, we point to the investigation of environmental influence on catalog completeness performed by \citet{Wainer22} as an example.

\subsection{Completeness Comparison: PHATTER vs.\ PHAT}

We find that the 50\% mass and luminosity completeness limits for the PHATTER M33 cluster catalog are worse relative to the similar PHAT M31 cluster catalog presented in \citetalias{Johnson15_AP}; fractional completeness is lower in M33 at a fixed mass or F475W magnitude.  At younger ages, the 50\% mass completeness limit for the M33 catalog is $\sim$0.3 dex higher than what was found for M31, and correspondingly $\sim$0.5 mags brighter in F475W magnitude.

While M33's larger distance can account for a 0.2 mag difference in luminosity, we believe the completeness differences are primarily due to the PHATTER footprint's central disk location and M33's higher star formation surface density, which together lead to a higher average density of young field stars stars (i.e., high $N_{\mathrm{MS}}$) and worse catalog completeness. Thanks to the use of an analogous $N_{\mathrm{MS}}$ definition by \citetalias{Johnson15_AP} for PHAT cluster work, we can confirm that the PHATTER median $N_{\mathrm{MS}}$ is a factor of 5 (0.7 dex) larger than the PHAT median $N_{\mathrm{MS}}$.  We also note that unlike for PHAT, using an alternative F475W$_{-3}$ metric, which subtracts the contribution of the three brightest stars from the integrated F475W magnitude, does not remove the age-dependent trend in luminosity completeness.

\section{Cluster SED Fitting} \label{sec:slugfitting}

In this section we discuss our method for deriving the cluster properties (age, mass, and extinction) from their integrated light photometry. We use the public source code Stochastically Lighting Up Galaxies (SLUG) \citep{Krumholz15_SLUG} to build a set of $10^7$ model star clusters that we use to estimate the M33 cluster properties in \S\ref{sec:slug_mod}. We discuss fitting of various filter combinations and discuss the reliability of this integrated light fitting in \S\ref{sec:slug_sel+fit} based on comparison to CMD-based results for a similar sample of M31 clusters presented in Appendix \ref{sec:app_slugM31}. We derive SLUG-based estimates for the cluster sample and present the results in \S\ref{sec:slug_result_m33}.  

\subsection{Building the SLUG Cluster Library} \label{sec:slug_mod}

SLUG is a stellar population synthesis code that incorporates stochastic modeling of stellar mass and luminosity distributions. More information and details about SLUG can be found in \citet{Krumholz15_SLUG}, and examples of its use include \citet{Krumholz15_LEGUS,Krumholz19_SLUG}. Using SLUG, we build a grid of $10^7$ model star clusters assuming Padova stellar evolution models that include thermally pulsing AGB stars \citep{Girardi00}, which are distributed with Starburst99 \citep{Vazquez05}. We use a \citet{Kroupa01} stellar initial mass function (IMF) which spans from 0.01 $M_\odot$ to 120 $M_\odot$ with the ``stop after'' sampling method, which allows for some of the more massive stars to be included in our simulated sample \citep{Krumholz15_SLUG}. We generate models with SLUG by drawing ages from a $t^{-1}$ distribution over a range of 10$^{6}$ to 10$^{10}$ years, encompassing the majority of clusters in the PHATTER catalog. We draw cluster masses from an $M^{-2}$ distribution and draw dust extinction values from a lognormal centered at an $A_V$ of 1 mag, width of 0.33 mag, and min/max breakpoints of 10$^{-6}$ and 5 mag. We apply extinction according to a Milky Way extinction curve \citep{Fitzpatrick99}, compute photometry in Vega magnitudes for six HST filters (ACS: F475W and F814W; WFC3: F275W, F336W, F110W, F160W), and convert from model magnitudes to observed magnitudes using an adopted distance modulus of 24.67 \citep{deGrijs14}.

\subsection{Filter Selection and Fitting} \label{sec:slug_sel+fit}

The reliability of SLUG cluster property determinations depends on the number of filters we are able to include in the fitted SEDs. Therefore, we begin by analyzing the fraction of clusters with good photometry in various combinations and numbers of filters for the PHATTER sample. Specifically, we choose filter combinations in increasing order of photometric detectability, and present the results for the M33 sample in Table~\ref{tab:M33_passband_percent}.

\vspace{-3.5em}

\begin{deluxetable}{cc}
\tabletypesize{\footnotesize}
\tablecaption{Detection Statistics for Passband Combinations \label{tab:M33_passband_percent}}
\tablehead{\colhead{Passband} & \colhead{N(Detections)}} 
\startdata
F275W+F336W+F475W+F814W+F110W+F160W  & 349  (28.7\%) \\
F336W+F475W+F814W+F110W+F160W  & 414 (34.1\%) \\
F275W+F336W+F475W+F814W  &  612 (50.4\%) \\
F336W+F475W+F814W & 729 (60.0\%)
\enddata
\tablecomments{The number and percentage of PHATTER clusters with photometric detections in each of the listed combinations of 3, 4, 5, and 6 filter passbands.}
\end{deluxetable}

\begin{deluxetable*}{ccc|ccc|ccc|ccc}
\tabletypesize{\footnotesize}
\tablecaption{SLUG Results \label{tab:slug_m33}}
\tablehead{
\colhead{ID} & \colhead{Error Cut Flag} & \colhead{Filters Available} & \multicolumn{3}{|c|}{log($Mass/M_{\odot}$)} & \multicolumn{3}{c|}{log($Age/yr$)} & \multicolumn{3}{c}{$A_{V}$}  \\
\colhead{} & \colhead{} & \colhead{} & 
\multicolumn{1}{|c}{P16} & \colhead{P50} & \colhead{P84} & \multicolumn{1}{|c}{P16} & \colhead{P50} & \colhead{P84} & \multicolumn{1}{|c}{P16} & \colhead{P50} & \colhead{P84}
}
\startdata
3 & F & 4 & 2.47 & 3.23 & 3.43 & 7.72 & 8.25 & 8.41 & 0.24 & 0.43 & 0.69 \\
5 & F & 6 & 3.99 & 4.07 & 4.16 & 8.41 & 8.49 & 8.57 & 0.23 & 0.41 & 0.57 \\
7 & F & 6 & 3.97 & 4.05 & 4.12 & 8.40 & 8.49 & 8.58 & 0.38 & 0.58 & 0.74 \\
8 & F & 6 & 4.53 & 4.59 & 4.65 & 8.17 & 8.23 & 8.28 & 0.20 & 0.26 & 0.32 \\
11 & F & 6 & 2.45 & 3.46 & 3.56 & 7.71 & 8.31 & 8.38 & 0.10 & 0.23 & 0.32
\enddata
\tablecomments{Table \ref{tab:slug_m33} is published in its entirety in the electronic edition of the {\it Astrophysical Journal}.  A portion is shown here for guidance regarding its form and content. The error cut flag identifies cases where fits have large uncertainties (16th to 84th percentile range $>$1.2~dex in age or $>$1.3~dex in mass) and should be excluded from uses where uncertainties are not factored in explicitly.}
\end{deluxetable*}

We compute SLUG fits for each filter combination listed in Table \ref{tab:M33_passband_percent} and the corresponding sample of clusters that are detected in all of the combination's selected passbands.  The SLUG model grid is trimmed for each passband combination to omit the magnitudes of any filters that are not selected.  Once the cluster fits from each filter combination are compiled, we adopt the fit for each cluster that results from the filter combination with the largest number of filters.  This ensures that we are not fitting incomplete SEDs and that we obtain fits for a maximum number of total clusters.

We execute each iteration of SLUG fitting, and process each set of results, using a fixed set of parameters and assumptions. We adopt the following settings that relate to specifics of the fitting process: a photometric bandwidth of 0.02, a physical properties bandwidth of 0.05, and a Gaussian kernel for PDF estimation. And same as for the underlying set of models, we assume a $t^{-1}$ age prior, $M^{-2}$ mass prior, and lognormal $A_V$ prior.  The code returns marginalized PDFs for age, mass, and dust extinction of each cluster, from which we can derive 16th, 50th, and 84th percentile values.  These percentiles yield median estimates for each physical property accompanied by an associated 1$\sigma$ uncertainty. We also flag and exclude highly uncertain fits, such that any cluster with a 16th to 84th percentile range greater than 1.2 dex in age or 1.3 dex in mass is identified by an error flag in the fitting results.

We present an example of a fitted cluster SED in the middle panel of Figure~\ref{fig:examplecluster} for PHATTER 22. The 100 best-fit SEDs from the SLUG library are plotted along with the observed SED, which show good agreement between models and observations. The median, 16th, and 84th percentiles of the marginalized posterior PDF for cluster age, mass, and dust extinction, computed over the full library of model SEDs, are derived using functions from the \texttt{cluster\_slug} package that is included as part of the SLUG code. We find good agreement between SED and CMD fitting results for PHATTER 22.

\subsection{SLUG Results} \label{sec:slug_result_m33}

We derive cluster properties using SLUG for 729 objects with detections in at least one of the filter combinations listed in Table~\ref{tab:M33_passband_percent}; we report the fitting results in Table \ref{tab:slug_m33}.  We note that the limited number of fitted clusters (729 out of 1214; 60\%) is due to a minimum three-filter (F336W, F475W, and F814W) detection criteria for SED fitting.  As a result, the completeness of this fitted sample of clusters is worse than the overall catalog completeness, and is biased toward younger and brighter clusters.

Excluding flagged cases with broad PDFs, the median ages from the cluster PDFs range from $6.08 < \mathrm{log(Age/yr)} < 8.91$ with a median value of 7.96 for our cluster sample.  The median cluster PDF masses range from $2.14 < \log(M/M_{\odot}) < 4.59$ with a median value of 3.29.  The median 16th to 84th percentile range in mass is 0.46~dex, and the median 16th to 84th percentile range in age is 0.41~dex.

To gauge the reliability of the SLUG fits for the PHATTER cluster sample, we compare newly-derived SLUG results to high-quality CMD-based cluster fits for a similar sample of clusters in M31.  We find that masses are reliably determined via SLUG integrated light fitting, but that SLUG age and dust results suffer from large uncertainties and artifacts.  In particular, SLUG fits tend not to reliably recover ages for clusters younger than $\sim$100~Myr, which instead are often fit with older ages; see Appendix~\ref{sec:app_slugM31} for full details of the comparison analysis and results.  Due to these results, we recommend that CMD-based fits for the younger clusters in the PHATTER cluster sample presented in \citet{Wainer22} should be preferred over the SLUG fits reported here.  At older ages, there are fewer resolved stars, making the SLUG age estimates the better (and sometimes the only) option.

\section{Discussion} \label{sec:discussion}

\subsection{Comparison to Existing M33 Cluster Catalogs}  \label{sec:litcat}

We cross-match the full PHATTER candidate list (clusters, galaxies, emission regions, and remaining ancillary objects; see Appendix~\ref{sec:app_othercat}) to five different catalogs from the literature: \citet{Sarajedini07}, \citet{SanRoman09}, \citet{SanRoman10}, \citet{Sharma11}, and \citet{Corbelli17}. These references are chosen to facilitate three types of comparisons: to comprehensive catalogs (\S\ref{sec:litcat_primary}), to HST-based catalogs (\S\ref{sec:litcat_hst}), and to infrared catalogs (\S\ref{sec:litcat_ir}). We compile the cross-matching results in Table~\ref{tab:compare} where we list identifiers (and object classes, where relevant) for each catalog, as well as additional alternate names and accompanying references.  These matches are based on a one arcsec match radius between cataloged positions, after applying a mean astrometric offset of 0$\farcs$609 to the \citet{Sarajedini07} cluster sample before cross-matching.

\begin{deluxetable*}{cccccccccc}
\centering
\tabletypesize{\footnotesize}
\setlength{\tabcolsep}{0.05in}
\tablecaption{Literature Cross-matching Results \label{tab:compare}}
\tablewidth{0pt}

\tablehead{
\colhead{ID} & \colhead{SM07 ID} & \colhead{SM07 Class\tablenotemark{a}} & \colhead{SR10 ID} & \colhead{SR10 Class\tablenotemark{b}} & \colhead{SR09 ID} & \colhead{S11 ID} & \colhead{C17 ID} & \colhead{C17 Class\tablenotemark{c}} & \colhead{Alternate Names \& References\tablenotemark{d}}}
\startdata
1 & \nodata & \nodata & 1849 & 3 & 157 & \nodata & \nodata & \nodata & \nodata \\
2 & 391 & Unknown & 2039 & 0 & \nodata & \nodata & \nodata & \nodata & CS U80 \\
3 & \nodata & \nodata & 2025 & 0 & \nodata & \nodata & \nodata & \nodata & \nodata \\
4 & \nodata & \nodata & 1441 & -1 & \nodata & \nodata & \nodata &\nodata  & \nodata \\
5 & 372 & Cluster & 1959 & 3 & \nodata & \nodata & \nodata & \nodata & CBF 58; MKKSS 50; CS U91
\enddata

\tablecomments{Table \ref{tab:compare} is published in its entirety in the electronic edition of the {\it Astrophysical Journal}.  A portion is shown here for guidance regarding its form and content. Literature references: SM07 \citep{Sarajedini07}, SR10 \citep{SanRoman10}, SR09 \citep{SanRoman09}, S11 \citep{Sharma11}, C17 \citep{Corbelli17}.}

\tablenotetext{a}{\citet{Sarajedini07} Classes: cluster, stellar, galaxy, unknown}
\tablenotetext{b}{\citet{SanRoman10} Classes: $-1$ = galaxy, 0 = unknown extended object, 1 = candidate cluster, 2 = highly probable cluster, and 3 = confirmed cluster}
\tablenotetext{c}{\citet{Corbelli17} Classes:   b = associated with clouds, no optical counterpart, c1 = associated with clouds: coincident $H_{\alpha}$ and mid-infrared  peaks, c2 = associated with clouds: coincident $H_{\alpha}$, mid-infrared and UV peaks, c3 = not associated with clouds but optically detected, d = ambiguous, e = not associated with clouds, mostly mid-infrared peaks only}
\tablenotetext{d}{Reference Abbreviations for Alternate Names \citep[from][]{Sarajedini07}: Hilt \citep{Hiltner60}, MD \citep{Melnick78}, CS \citep{Christian82}, MKKSS \citep{Mochejska98}, CBF \citep{Chandar99, Chandar01}, BEA \citep{Bedin05}, SBGHS \citep{SarajediniBarker07}}
\end{deluxetable*}
\vspace{-2em}


\begin{figure*}
    \centering
    \includegraphics[width=\textwidth]{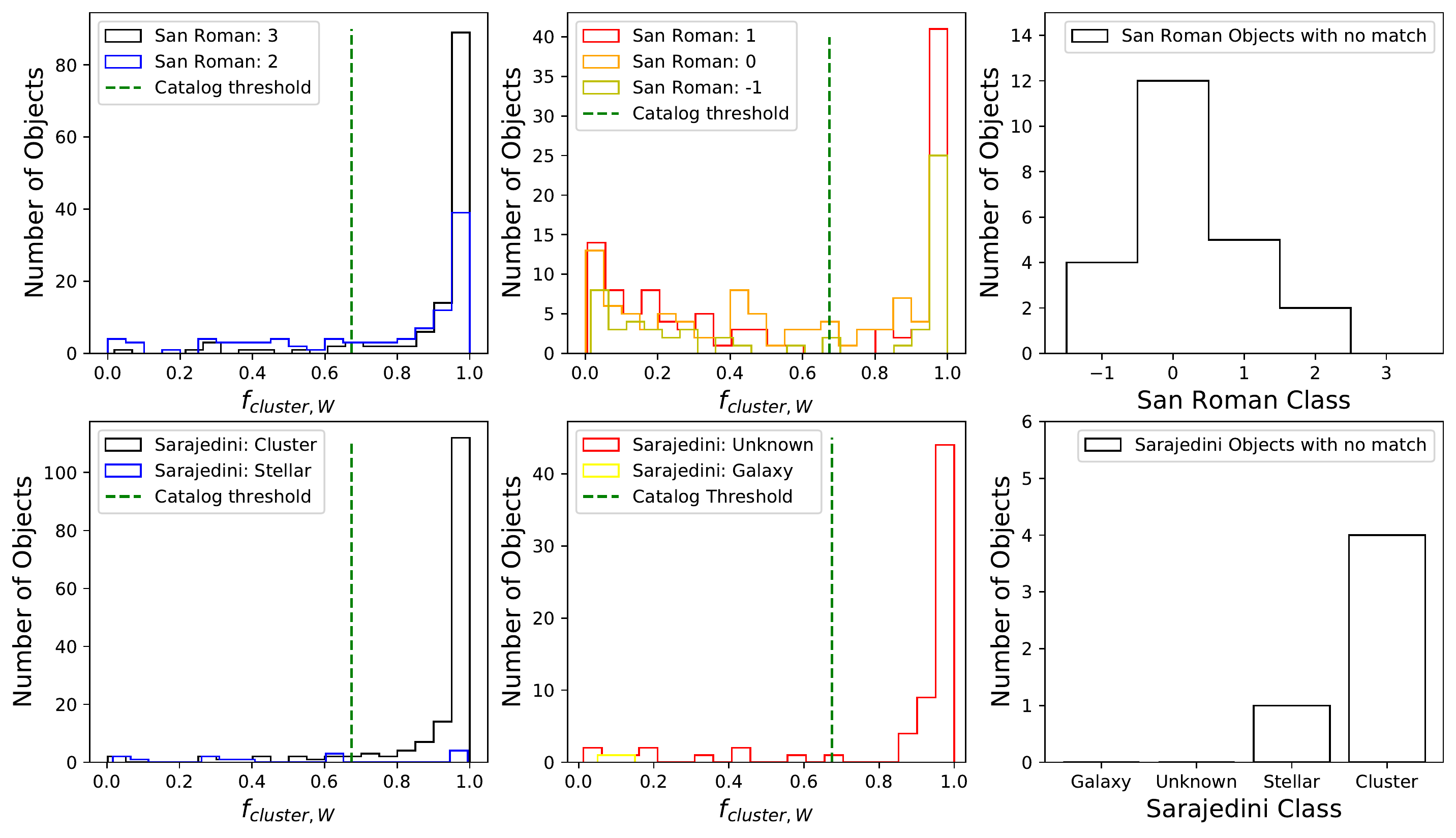}
    \caption{A comparison of LGCS classifications for objects in the \citet{SanRoman10} and \citet{Sarajedini07} catalogs that fall within the PHATTER footprint. Top: Left and center panels show $f_{\mathrm{cluster,W}}$ distributions for \citet{SanRoman10} catalog, where object classes include ``confirmed star cluster'' (class 3), ``highly probable star cluster'' (class 2), ``candidate star cluster'' (class 1), ``unknown'' (class 0), and ``background galaxy'' (class $-1$). The dashed line represents the $f_{\mathrm{cluster,W}}$ threshold we use for cluster catalog inclusion. The right panel shows the \citet{SanRoman10} classes for objects located in the PHATTER footprint that did not match to a LGCS identification. Bottom: same as top, but for matched (left and center) and unmatched (right) objects from the \citet{Sarajedini07} catalog and its different set of object classes.}
    \label{fig:catalogcompare}
\end{figure*}


\subsubsection{Primary Catalog Comparisons: Sarajedini \& Mancone (2007) and San Roman et al.\ (2010)} \label{sec:litcat_primary}

We compare the PHATTER cluster catalog to two key M33 cluster catalogs in the literature: \citet{Sarajedini07} and \citet{SanRoman10}.  We focus on these catalogs for our primary literature comparison due to their comprehensive compilation of published M33 cluster catalogs and their complete, uniform spatial coverage, respectively.

We summarize the recovery of clusters from the optically-selected catalogs of \citet{Sarajedini07} and \citet{SanRoman10} in Figure~\ref{fig:catalogcompare}.  We recover a large fraction of the cataloged cluster candidates that fall within the spatial coverage of the PHATTER survey. Specifically, for objects that \citet{Sarajedini07} classify as ``cluster'' and lie within the PHATTER footprint, 89\% are present in the PHATTER cluster catalog. We also classify 85\% of their ``unknown'' objects as clusters. 

For the portion of the \citet{SanRoman10} catalog that falls within the PHATTER footprint, 92\% of their ``confirmed star cluster'' (class 3) objects and 67\% of their ``highly probable star cluster'' (class 2) objects are present in the PHATTER cluster catalog.  Most of the remaining class 2 and 3 \citet{SanRoman10} objects are recovered by the LGCS search, but lie in a long tail at low $f_{\mathrm{cluster,W}}$ values, as shown in the top left panel of Figure~\ref{fig:catalogcompare}. Additionally, 41\% of their less certain cluster identifications (``candidate star cluster'' / class 1 and ``unknown'' / class 0) are also identified as clusters in this work. More surprisingly, we find that half of the ``background galaxy'' (class $-1$) objects that lie within the PHATTER footprint are classified as clusters by the LGCS search. Visual inspection of HST images for these objects confirm they are in fact clusters, and thus were misclassified by the ground-based \citet{SanRoman10} effort.

Finally, we note that a small fraction of \citet{Sarajedini07} and \citet{SanRoman10} sources that fall within the PHATTER footprint were not recovered in the LGCS search. We show the class distribution of these objects in the right panels of Figure~\ref{fig:catalogcompare}.  

\subsubsection{Space-based Comparisons: HST Catalogs}  \label{sec:litcat_hst}

We compare the PHATTER cluster catalog to three HST-based cluster catalogs from the literature: \citet[][hereafter, collectively CBF]{Chandar99, Chandar01}, \citet{Bedin05}, and \citet{SanRoman09}.  We note that CBF and \citet{Bedin05} were cross-matched as members of the \citet{Sarajedini07} compilation, and \citet{SanRoman09} was matched individually.  While these literature catalogs were derived from imaging datasets with relatively small spatial footprints, they serve as useful points of comparison for analyzing catalog-specific differences in visual cluster identification of HST imagery.

A comparison between the PHATTER catalog and the CBF catalog shows that the PHATTER catalog recovers nearly all previously identified objects (103 out of 110 that lie within PHATTER footprint).  However, the PHATTER catalog includes an additional 445 clusters within the spatial footprint searched by CBF, resulting in a total catalog that is a factor of 5 larger. This discrepancy is expected due to CBF's use of WFPC2 images whose wide field cameras have significantly lower spatial resolution (pixel scale of 0.1 arcsec) than the ACS and WFC3 instruments used by the PHATTER survey (pixel scales of 0.05 and 0.04 arcsec, respectively). Lower spatial resolution images make the identification of faint, low mass clusters much more difficult, leading to worse luminosity and mass completeness limits for the CBF catalog and significantly fewer cluster identifications as a result.
    
Next we compare the PHATTER catalog to the work of \citet{Bedin05}, who used a single ACS pointing located within the PHATTER survey footprint to identify 33 clusters. Within this spatial region, the PHATTER catalog includes 33 clusters, where 22 entries are shared between the two works.  The 11 unmatched clusters from the \citet{Bedin05} catalog break down into two categories: (1) 6 objects are explained by object definition differences, where PHATTER categorized these objects as emission regions or loose non-cluster associations; (2) 5 objects are identified by the PHATTER search, but are excluded from the cluster catalog due to low $f_{\mathrm{cluster, W}}$.  For the 11 PHATTER clusters not recovered by \citet{Bedin05}, we believe the mismatch is due to their use of a single-band F775W ACS image.  These clusters have lower $f_{\mathrm{cluster, W}}$ and fainter $m_{\mathrm{F475W}}$ than most of the 22 matched clusters.  Faint, low mass clusters are identified in PHATTER via a small clustering of blue main sequence stars, and therefore it is expected that these objects would be missed in a search of only red wavelength imagery.
    
Finally, we compare the PHATTER catalog to the work of \citet{SanRoman09}, who searched multi-band ACS imagery that partially overlaps with the PHATTER survey footprint and identified 86 clusters in the overlapping region. The PHATTER cluster catalog includes 75 of these previously identified objects, leaving 11 objects that were excluded by the PHATTER catalog selection threshold.  Importantly, the PHATTER catalog includes 119 clusters not identified by \citet{SanRoman09}, resulting in a total catalog that is a factor of 2.3 larger.  Upon examination, this significant difference is due to a more conservative selection threshold, where the PHATTER catalog tends to probe to lower $f_{\mathrm{cluster, W}}$ and fainter $m_{\mathrm{F475W}}$. Imagery wavelength may also play a role here, as more than half of the PHATTER overlapping fields were only imaged in two redder bands (F606W and F814W) without bluer F475W coverage.

\begin{figure*}
    \centering
    \includegraphics[width=0.65\textwidth]{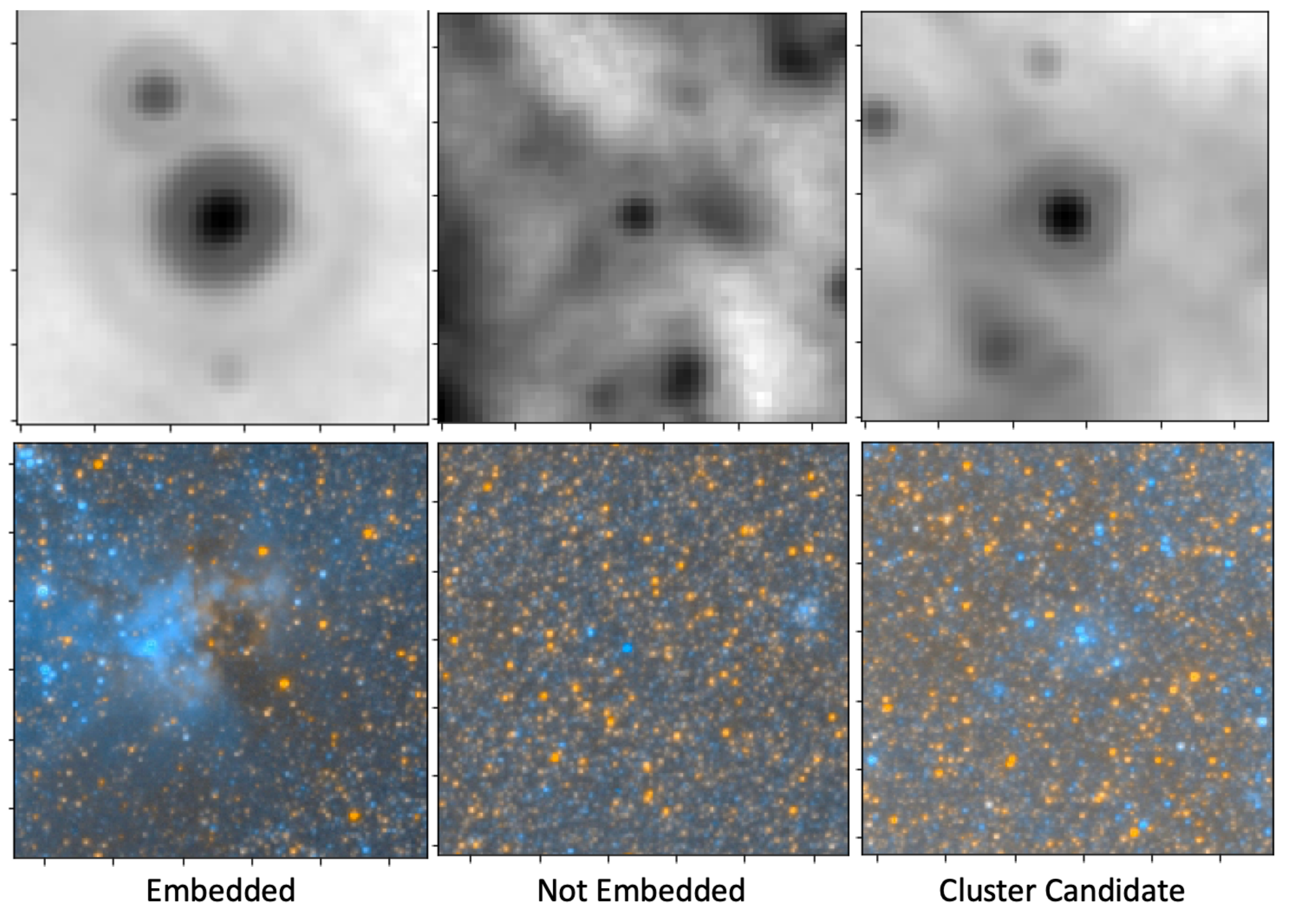}
    \caption{Spatially-matched Spitzer 24$\mu$m (top) and HST F475W$-$F814W (bottom) color cutouts for three \citet{Corbelli17} and \citet{Sharma11} objects that represent three scenarios for the objects: an object which shows evidence of being a young embedded star forming region (left; Corbelli YSCC 225), an object where there is sufficient background visible to rule out an embedded cluster (middle; Corbelli YSCC 215), and a rare object which we classify as a star cluster in the PHATTER catalog (right; Corbelli YSCC 297, PHATTER 529).}
    \label{fig:c_o}
\end{figure*}


\subsubsection{Infrared Comparisons: Alternative Catalogs}  \label{sec:litcat_ir}

Studies in the literature have also made use of infrared images to assess M33's cluster population. \citet{Sharma11} uses Spitzer 24 $\mu$m images to construct a catalog of young stellar clusters, where it is assumed these objects are still embedded in their natal molecular clouds. Of the 240 \citet{Sharma11} objects that fall within the PHATTER footprint, we only identify 41 cross-matches, of which we only classify 5 as star clusters while the remainder are mostly low-$f_{\mathrm{view}}$ candidates. The poor correspondence between these two catalogs corroborates the conclusions of \citet{Sun16}, who advocate against the use of the \citet{Sharma11} catalog for star cluster population studies due to contamination by non-cluster objects, as we noted in \S\ref{sec:intro}.

The catalog of \citet{Sharma11} was used as a starting point for compiling a sample of young cluster candidates for use in a cross-comparison with a CO molecular cloud catalog by \citet{Corbelli17}. Of the 291 objects from \citet{Corbelli17} within the PHATTER footprint, we match 91 objects and find just 12 to be star clusters identified in our catalog. As with the \citet{Sharma11} sample, these cluster candidates are generally not associated with optical star clusters, and caution should be taken when using this catalog for the purpose of star cluster population work.

To provide a visual example of optical vs.\ infrared cluster candidates, we present Spitzer 24$\mu$m and HST F475W+F814W color cutout pairs for three \citet{Corbelli17} and \citet{Sharma11} objects in Figure~\ref{fig:c_o}. The first is an example of a region where there is a bright 24$\mu$m source, and in the color image there is visible extinction, indicative of molecular gas, and an emission region.  These indicators confirm the presence of a young embedded star forming region and potentially (but not certainly) a young star cluster.  In the absence of certainty on the presence of a bound star cluster, this object is not identified as a cluster in the PHATTER catalog. The second example is an object where, even though there is a 24$\mu$m source, the background field in the optical image is fully visible with no sign of dust obscuration. Therefore, we conclude the probability of an embedded cluster is low. The third example is an object that is a classified as a cluster in our catalog. Examination of PHATTER imaging around these positions suggests that for at least 30\% of \citet{Corbelli17} objects, we can rule out the presence of an embedded star cluster based on the uniformity of the background at the location of the \citet{Corbelli17} cluster candidates.  Thus it appears that a significant fraction of these candidates are neither young embedded star forming regions nor optically visible star clusters.

\subsubsection{Catalog Comparison Summary}  \label{sec:litcat_summary}

\begin{figure}
    \centering
    \includegraphics[width=0.47\textwidth]{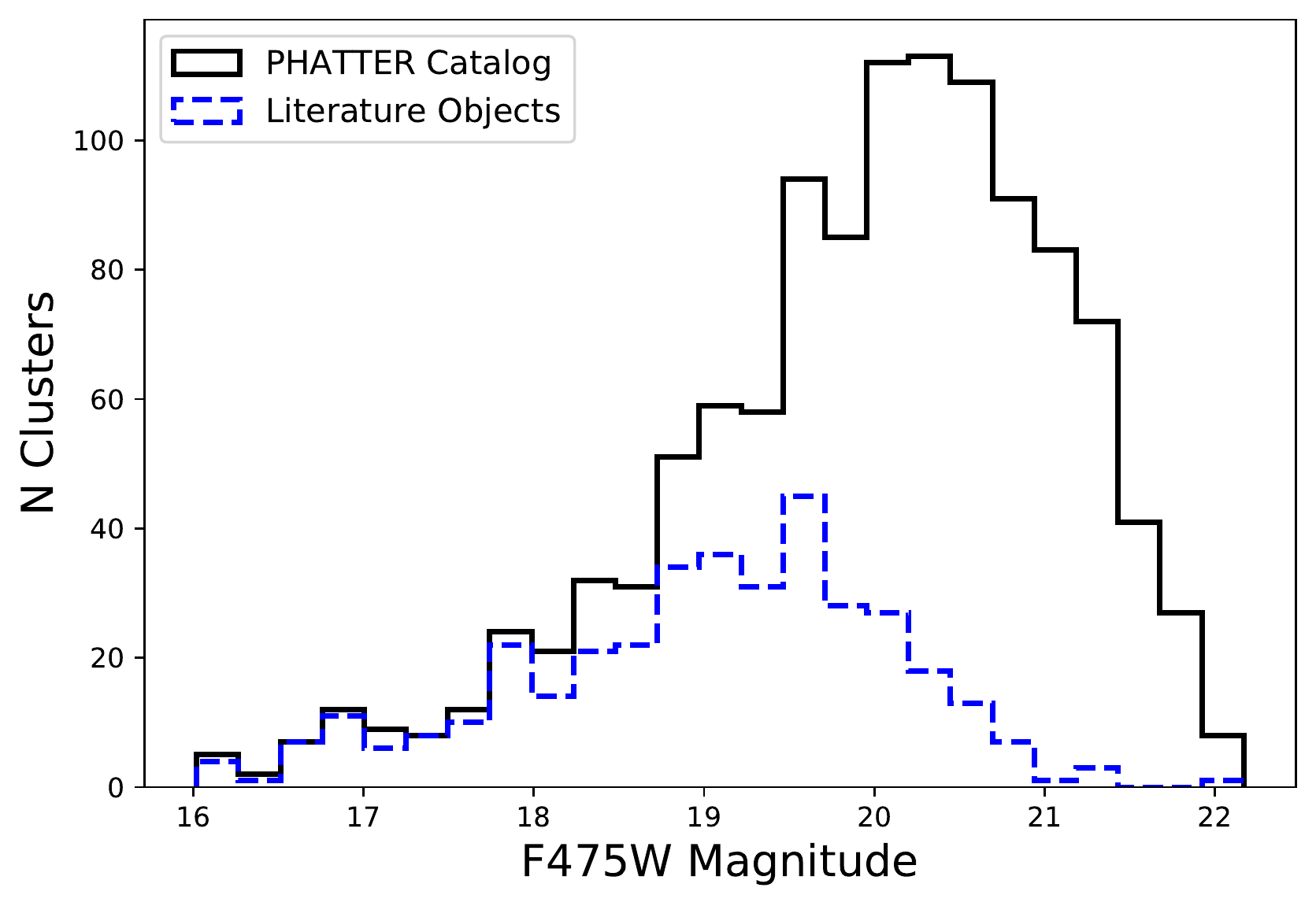}
    \caption{Shown here are the F475W magnitude distributions of PHATTER clusters (black), and previously identified literature objects (blue). Shown in black are 1166 PHATTER clusters with F475W detections. In blue are 391 literature objects included in the PHATTER cluster catalog with F475W detections. Literature objects not represented above: 13 objects lacking F475W detections and 178 objects with low $f_{cluster,W}$ omitted from the PHATTER cluster catalog.}
    \label{fig:lum_comp}
\end{figure}

Overall, the PHATTER cluster catalog significantly enhances the population of known clusters in the inner disk region of M33. Out of 1214 total clusters, 810 (67\% of the catalog) are identified here for the first time. We compare the luminosity distributions of PHATTER clusters and previously identified objects from the literature in Figure \ref{fig:lum_comp}, showing that PHATTER's sample probes fainter, lower mass clusters than previous works. The superior spatial resolution of HST images facilitate this marked increase, as tight groups of stars can be differentiated from a single, barely or unresolved source.  In addition, the ability to identify small groupings of faint resolved stars in PHATTER images leads to our ability to probe further down the cluster mass function in M33 than ever before.

\subsection{M33 Catalog in Context: Expectations from \texorpdfstring{$\Gamma$}{Gamma} and \texorpdfstring{$M_c$}{Mc} Correlations} \label{sec:m33_sample_predict}

We examine whether the PHATTER sample size of 1214 clusters matches expectations based on our knowledge of M33 and the relation between galaxies and their cluster populations.  An easy comparison we can make is to the PHAT M31 catalog \citepalias{Johnson15_AP} and its 2753 clusters. We might naively expect M31's sample to outnumber M33's based on the number of bricks in the survey footprint (23 vs.\ 3) or the total galaxy stellar mass \citep[6$\times$10$^{10}$ $M_{\odot}$ vs.\ 3$\times$10$^{9}$ $M_{\odot}$;][]{Patel17}.
In this context, it seems there should be a much larger than factor of $\sim$2 difference in cluster count.  However, differences in on-going star formation and spatial coverage fraction of HST imaging could plausibly provide an explanation, especially given that both cluster samples are dominated by young star clusters. 
Therefore, in this section we endeavor to make an SFR-based accounting of the M33 cluster sample to test whether we can explain the sample size differences we see between M33 and M31.  We will use observed relations with respect to galaxy \SigSFR\ to predict cluster formation efficiency ($\Gamma$) and Schecter trunctation mass ($M_c$), and use these inputs to infer a prediction for cluster sample size.

To begin, we derive a SFR and \SigSFR\ measurement for the PHATTER survey region in M33.  Following the methodology laid out in Appendix A of \citet{Johnson17}, we use \textit{GALEX} FUV and \textit{Spitzer} 24 \textit{$\mu$m} images and a \SigSFR\ prescription from \citet{Leroy08}. We find log(\CSigSFR\ / \solperyr\ kpc$^{-2}$) of $-2.04^{+0.16}_{-0.18}$ for M33 within the PHATTER survey footprint.  We also extract the total star formation rate within the survey region, which we find is 0.12 \solperyr, approximately 46\% of M33's total SFR.

Next, we can use the PHATTER \SigSFR\ value to derive an estimate of the cluster formation efficiency, $\Gamma$, based on the $\Gamma$-\SigSFR\ relation from \citet{Johnson16_gamma}.  This relation predicts a cluster formation efficiency of 15\%.  We can also derive an estimate for the cluster mass function by assuming a Schechter function form with a $-2$ power law slope and trunctation mass, $M_c$ estimate based on the $M_c$-\SigSFR\ relation from \citet{Johnson17}.  This relation predicts a Schechter $M_c$ of 4.3$\times$10$^4$ \Msun. 

When we combine the determinations of M33 SFR and $\Gamma$, we derive a cluster formation rate (CFR) of 0.0180 \solperyr.  Interestingly, this is very similar to the PHAT M31 CFR of 0.0186 ($\Gamma$=6.4\% via \citealt{Johnson16_gamma} and SFR=0.29 via \citealt{Lewis15}). Although the PHATTER M33 SFR is 2.4 times smaller then the PHAT M31 survey SFR (0.12 \solperyr vs.\ 0.29 \solperyr), M33's \SigSFR\ is almost 4 times higher, which leads to a 2.4x higher $\Gamma$ (15\% vs.\ 6.3\%) that cancels out the SFR difference.

Based on the well-matched M31 and M33 CFRs, the relative difference in size between the two cluster catalogs (2753 vs.\ 1214; 2.3x) seems unexpected.  Note that the variation in $M_c$ (8.5$\times$10$^3$ \Msun\ vs.\ 4.3$\times$10$^4$ \Msun) is only expected to make a few percent difference in the number statistics due to the small number of clusters at the high mass end, so that does not explain the difference. However, catalog completeness differences likely play a role. M33's high stellar density within the PHATTER footprint in the central region of the disk leads to a higher mass for the 50\% completeness limit than in M31: log($M$/\Msun) of 3.2 vs.\ 3.0 for a nominal 100--300 Myr age range. However, this 0.2 dex offset only affords a 1.3x--1.7x correction to M33 $N_{\mathrm{cluster}}$, leaving another factor of 1.3x--1.7x still unexplained.  This remaining discrepancy could be due to our assumption of a constant SFH, or perhaps we will find that $\Gamma$ is not as high in M33 as predicted.

Overall, this exercise shows that the star formation differences between M31 and M33 are likely significant enough to impact cluster populations.  We look forward to using robust star formation history fitting \citep{Lazzarini22} to inform recent SFR and \SigSFR\ determinations for M33, and aid future $\Gamma$ determinations for the PHATTER cluster sample.

\subsection{Cluster Affiliated Phenomena: X-ray Sources \& Planetary Nebulae} \label{sec:xraypn}

One of the immediate uses of the PHATTER cluster catalog is to cross-match it with objects of interest in M33, such as planetary nebulae (PNe), X-ray emitting sources, and other stellar populations. Identifying associations between clusters and these source populations can provide useful information about the source, such as a cluster-based age or additional information that assists in source classification.

We begin by searching for cross-matches between the PHATTER cluster catalog and two M33 X-ray source catalogs created from Chandra \citep[ChASeM33 survey;][]{Tullmann11} and XMM-Newton \citep{Williams15_xray} observations.  We use an initial matching radius of 5 arcsec, but require the X-ray source to fall within the aperture radius of the cluster center (typically $\sim$1.5 arcsec), resulting in two matches --- one from each X-ray catalog. The matched Chandra source (ChASeM33 393) is paired with PHATTER 675, a known globular cluster \citep[ID: 275;][]{Sarajedini07} whose association with an X-ray source was identified by \citet{Tullmann11}.  The matched XMM-Newton catalog entry (Source 716) is associated with PHATTER 29, a previously identified young cluster \citep[ID: 260;][]{Sarajedini07} with a CMD-estimated age of 10 Myr \citep{Wainer22}.  Given the ages of the clusters, the Chandra source in the globular cluster is likely to be a bright low-mass X-ray binary, and the XMM-Newton source in the young cluster is likely to be a bright high-mass X-ray binary.

We also report the presence of a possible planetary nebula (PN) associated with a PHATTER cluster.  The candidate PN associated with the PHATTER 4 cluster was discovered serendipitously while reviewing sources with outlier optical colors in cluster CMDs. PNe are known to appear as anomalous blue sources in F475W$-$F814W color due to strong line emission in the F475W bandpass \citep{Veyette14}. PHATTER 4 has a very uncertain SLUG integrated light age determination, most likely due to the unmodeled contribution of nebular line emission from the PN, however its CMD-fitted age estimate of $\sim$1 Gyr \citep{Wainer22} is not unexpected for a PN-hosting star cluster.

Given the serendipitous identification of the PN, we conducted a search for additional candidates.  We performed a cross-match of the PHATTER clusters with the PN catalog of \citet{Ciardullo04}, but we found no candidates that lie within the aperture radius from a cluster center.  This lack of matches is not unexpected, however, given that a cluster PN would likely have been rejected by the \citet{Ciardullo04} search of groundbased narrowband imaging due to the presence of coincident continuum emission from the cluster.  We also perform a search for other cluster members with anomalous optical colors (F475W$-$F814W $<$ $-1$), but find no other reliable sources among cluster members.

We will continue to expand PHATTER cluster catalog cross-matching and the analyses of cluster membership to additional source populations in future work.  Following on from work conducted for PHAT, cluster membership of AGB stars \citep{Girardi20}, Cepheid variables \citep{Senchyna15}, and other populations are ripe for study in the PHATTER data.

\section{Summary} \label{sec:summary}

We present the results of a crowdsourced visual star cluster search of M33 conducted as part of the LGCS citizen science project using imaging from the PHATTER survey.  The resulting catalog of 1214 star clusters has well-characterized completeness properties and a 50\% completeness limit of approximately 1500 \Msun\ at an age of 100 Myr. We derive ages and masses from SED fitting of the subset of clusters with multi-band detections in the catalog's integrated aperture photometry. We find the sample is composed primarily of young, low-mass star clusters, although the SLUG-fitted clusters are a biased subsample of the full PHATTER catalog.

This cluster catalog builds upon similar Local Group cluster work in M31 \citepalias{Johnson15_AP} and significantly increases the number of known Local Group star clusters observed with HST. The PHATTER cluster catalog samples higher \SigSFR\ galactic properties than M31, which provides leverage for studying how cluster properties like the cluster mass function, cluster formation efficiency, and more depend on star formation intensity.  In accompanying work, we use CMDs of individually resolved stars to fit high-precision ages and masses, and to constrain the mass function of young clusters \citep{Wainer22}. We also expect the sample will also be useful for calibrating models of stellar evolution \citep[e.g.,][]{Girardi20} and other future M33 cluster studies.


\begin{acknowledgements}
We enthusiastically thank the $\sim$2,800 LGCS volunteers who made this work possible. Their contributions are acknowledged individually at \url{https://authors.clustersearch.org}. This publication uses data generated via the Zooniverse.org platform, development of which is funded by generous support, including a Global Impact Award from Google, and by a grant from the Alfred P. Sloan Foundation. Support for this work was provided by NASA through grant number HST-GO-14610 from the Space Telescope Science Institute, which is operated by AURA, Inc., under NASA contract NAS5-26555. L.C.J. acknowledges support through a CIERA Postdoctoral Fellowship at Northwestern University. This material is partially based on work by T.M.W. as a CIERA REU student at Northwestern University, supported by the National Science Foundation under grant No. AST-1757792. Some of the data presented in this paper were obtained from the Mikulski Archive for Space Telescopes (MAST).  This research made use of NASA's Astrophysics Data System (ADS) bibliographic services.
\end{acknowledgements}

\facilities{HST(ACS, WFC3)}
\software{astropy \citep{Astropy13}, DOLPHOT \citep{Dolphin00}, DrizzlePac \citep{DrizzlePac12,Hack13,Avila15}, SLUG \citep{Krumholz15_SLUG}, TOPCAT \citep{TOPCAT05}}


\appendix

\section{Ancillary Catalogs} \label{sec:app_othercat}

To accompany the primary PHATTER cluster catalog described in Section \ref{sec:catalog} and presented in Table \ref{tab:cat}, we present ancillary catalog results in this Appendix.  First, we present an ancillary catalog in Table \ref{tab:anc_cat}, which contains 3566 candidate identifications with $f_{\mathrm{cluster,W}} < 0.674$ and $f_{\mathrm{view}} \ge 0.1$. These objects fall below the catalog threshold adopted for the primary PHATTER cluster sample, however we publish these results to enable investigators to make their own decisions regarding completeness vs.\ contamination, depending on their science case.

Second, we present object-by-object synthetic cluster results in Table \ref{tab:syncat}. These results can be used to derive catalog completeness for the case where a different catalog selection is adopted.  

Third, we identify a sample of background galaxies using a selection criteria of $f_{\mathrm{galaxy}} \ge 0.25$.  We present this sample of 203 background galaxies identified by LGCS participants in Table \ref{tab:bckgal}.  The adopted $f_{\mathrm{galaxy}}$ detection threshold was chosen via visual inspection and ensures good sample purity. We find 7 cross-matches with \citet{SanRoman10}, 3 cross-matches with \citet{Sarajedini07}, and 3 cross-matches with \citet{Corbelli17} for the background galaxy sample.

Fourth, we identify a sample of emission regions using a selection criteria of $f_{\mathrm{emission}} \ge 0.25$.  We present this sample of 95 emission regions identified by LGCS participants in Table \ref{tab:emissionreg}.  The adopted $f_{\mathrm{emission}}$ detection threshold was chosen via visual inspection and ensures good sample purity.  We urge caution when interpreting this sample due to the fact that cluster identification, not emission region identification, was the primary goal of the LGCS search.  As a result, we do not expect this sample to be systematically complete, and instead recommend that these objects be considered opportunistic identifications. We note that three objects listed here are also included in the PHATTER cluster catalog, but where we agree with the cross-listing between the two classification categories: objects 163, 275, and 652. Also, we find 41 cross-matches with \citet{SanRoman10}, 15 cross-matches with \citet{Sarajedini07}, and 21 cross-matches with \citet{Corbelli17} for the emission region sample.

\begin{deluxetable*}{cccccccccccccccccccccccc}
\tabletypesize{\tiny}
\setlength{\tabcolsep}{0.05in}
\tablecaption{Ancillary Catalog \label{tab:anc_cat}}
\tablewidth{0pt}

\tablehead{
\colhead{ID} & \colhead{RA (J2000)} & \colhead{DEC (J2000)} & \colhead{$R_{\mathrm{ap}}$ ($\arcsec$)} & \colhead{$R_{\mathrm{eff}}$ ($\arcsec$)} & \colhead{$f_{\mathrm{view}}$} & \colhead{$f_{\mathrm{cluster}}$} & \colhead{$f_{\mathrm{galaxy}}$} & \colhead{$f_{\mathrm{emission}}$} & \colhead{$f_{\mathrm{cluster, W}}$} & \colhead{Flags} & \colhead{$m_{\mathrm{apcor}}$} & \colhead{$m_{275}$} & \colhead{$\sigma_{275}$} & \colhead{$m_{336}$} & \colhead{$\sigma_{336}$} & \colhead{$m_{475}$ }& \colhead{$\sigma_{475}$} & \colhead{$m_{814}$} & \colhead{$\sigma_{814}$} & \colhead{$m_{110}$} & \colhead{$\sigma_{110}$} & \colhead{$m_{160}$} & \colhead{$\sigma_{160}$}
}

\startdata
37 & 23.474286 & 30.498702 &  2.10 &  0.89 & 0.9333 & 0.3583 & 0.4667 & 0.1083 & 0.4833 & \nodata & -0.31 & \nodata & \nodata & \nodata & \nodata &  20.48 &   0.12 &  18.86 &   0.01 & \nodata & \nodata & \nodata & \nodata \\
46 & 23.363899 & 30.699947 &  1.90 &  0.35 & 0.9667 & 0.3000 & 0.5833 & 0.0833 & 0.3108 & \nodata & -0.01 & 19.49 &   0.21 &  19.28 &   0.13 &  19.23 &   0.06 &  17.69 &   0.08 & \nodata & \nodata & \nodata & \nodata \\
54 & 23.400054 & 30.709211 &  2.79 &  0.69 & 0.9667 & 0.2667 & 0.0000 & 0.7000 & 0.3235 & \nodata & -0.06 & 17.33 &   0.14 &  17.41 &   0.07 &  18.73 &   0.11 &  18.56 &   0.14 &  17.93 &   0.12 &  $>$18.28 & \nodata \\
57 & 23.469856 & 30.730419 &  3.84 &  1.48 & 0.9583 & 0.3583 & 0.0000 & 0.6000 & 0.4373 & \nodata & -0.25 & 16.49 &   0.07 &  16.60 &   0.11 &  17.62 &   0.12 &  17.24 &   0.10 & \nodata & \nodata & \nodata & \nodata \\
60 & 23.527783 & 30.815464 &  3.74 &  0.81 & 0.9667 & 0.2500 & 0.0000 & 0.7167 & 0.2552 & \nodata & -0.03 & 16.11 &   0.02 &  16.34 &   0.05 &  17.48 &   0.02 &  17.56 &   0.08 &  17.13 &   0.06 &  $>$17.88 & \nodata
\enddata
\tablecomments{Table \ref{tab:anc_cat} is published in its entirety in the electronic edition of the {\it Astrophysical Journal}.  A portion is shown here for guidance regarding its form and content.}
\end{deluxetable*}
\vspace{-2em}

\begin{deluxetable*}{cccccccccccccccccc}
\tabletypesize{\scriptsize}
\setlength{\tabcolsep}{0.05in}
\tablecaption{Synthetic Cluster Results \label{tab:syncat}}
\tablewidth{0pt}

\tablehead{\colhead{SYNID} & \colhead{RA (J2000)} & \colhead{DEC (J2000)} & \colhead{log($M/M_{\odot}$)} & \colhead{log(Age/yr)} & \colhead{$Z$} & \colhead{$A_V$} & \colhead{$R_{\mathrm{eff, in}}$ (pc)} & \colhead{$F475W_{\mathrm{in}}$} & \colhead{$F814W_{\mathrm{in}}$} & \colhead{$f_{\mathrm{view}}$} & \colhead{$f_{\mathrm{cluster}}$} & \colhead{$f_{\mathrm{galaxy}}$} & \colhead{$f_{\mathrm{emission}}$} & \colhead{$f_{\mathrm{cluster, W}}$} & \colhead{$N_{\mathrm{MS}}$} & \colhead{$N_{\mathrm{RGB}}$} & \colhead{Detected}}
\startdata
1 & 23.638290 & 30.781798 & 2.79 & 7.50 & 0.0152 & 0.44 & 1.09 & 20.64 & 20.59 & 0.0 & 0.0 & 0.0 & 0.0 & 0.0 & 3184 & 163 & False \\
2 & 23.630015 & 30.783396 & 2.86 & 7.35 & 0.0152 & 0.28 & 7.62 & 19.93 & 20.04 & 0.0 & 0.0 & 0.0 & 0.0 & 0.0 & 2186 & 180 & False \\
3 & 23.620705 & 30.785866 & 3.25 & 6.85 & 0.0152 & 2.75 & 9.64 & 20.66 & 19.43 & 0.05 & 0.05 & 0.0 & 0.0 & 0.0397 & 778 & 191 & False \\
4 & 23.629348 & 30.790771 & 2.43 & 6.85 & 0.0152 & 0.15 & 1.17 & 18.80 & 15.64 & 0.1833 & 0.1167 & 0.0167 & 0.05 & 0.0478 & 839 & 120 & False \\
5 & 23.624151 & 30.789617 & 3.22 & 8.95 & 0.0152 & 0.85 & 2.09 & 22.44 & 21.27 & 0.1475 & 0.1475 & 0.0 & 0.0 & 0.1740 & 676 & 167 & False
\enddata
\tablecomments{Table \ref{tab:syncat} is published in its entirety in the electronic edition of the {\it Astrophysical Journal}. A portion is shown here for guidance regarding its form and content. Synthetic clusters 1--848 are the first batch of randomly distributed tests, and clusters 849--1696 are the second batch of tests spatially distributed to place young clusters in regions with higher $N_{\mathrm{MS}}$. The boolean ``detected'' column reflects whether the synthetic cluster is selected by the final cluster catalog selection criteria: $f_{\mathrm{cluster,W}} > 0.674$.}
\end{deluxetable*}
\vspace{-2em}

\begin{deluxetable*}{ccccccccccc}
\tabletypesize{\scriptsize}
\setlength{\tabcolsep}{0.05in}
\tablecaption{Background Galaxy Catalog \label{tab:bckgal}}
\tablewidth{0pt}

\tablehead{
\colhead{ID}& \colhead{RA (J2000)} & \colhead{DEC (J2000)} & \colhead{$R_{\mathrm{ap}}$ ($\arcsec$)} & \colhead{$f_{\mathrm{view}}$} & \colhead{$f_{\mathrm{cluster}}$} & \colhead{$f_{\mathrm{galaxy}}$} & \colhead{$f_{\mathrm{emission}}$} & \colhead{$f_{\mathrm{cluster, W}}$} & \colhead{$m_{814}$} & \colhead{$\sigma_{814}$}
}

\startdata
24 & 23.647657 & 30.805648 & 6.40 & 0.9833 & 0.1000 & 0.8833 & 0.0000 & 0.0855 & 18.03 & 0.07\\
37 & 23.474286 & 30.498702 & 2.10 & 0.9333 & 0.3583 & 0.4667 & 0.1083 & 0.4833 & 18.86 & 0.01\\
46 & 23.363899 & 30.699947 & 1.90 & 0.9667 & 0.3000 & 0.5833 & 0.0833 & 0.3108 & 17.69 & 0.08\\
108 & 23.557170 & 30.482706 & 2.27 & 0.9500 & 0.0833 & 0.8667 & 0.0000 & 0.1116 & 20.16 & 0.37\\
126 & 23.484892 & 30.491225 & 2.64 & 0.9344 & 0.0656 & 0.8525 & 0.0164 & 0.1497 & 17.00 & 0.05
\enddata
\tablecomments{Table \ref{tab:bckgal} is published in its entirety in the electronic edition of the {\it Astrophysical Journal}.  A portion is shown here for guidance regarding its form and content.}
\end{deluxetable*}
\vspace{-2em}

\begin{deluxetable*}{ccccccccc}
\tabletypesize{\scriptsize}
\setlength{\tabcolsep}{0.05in}
\tablecaption{Emission Region Catalog \label{tab:emissionreg}}
\tablewidth{0pt}

\tablehead{
\colhead{ID}& \colhead{RA (J2000)} & \colhead{DEC (J2000)} & \colhead{$R_{\mathrm{ap}}$ ($\arcsec$)} & \colhead{$f_{\mathrm{view}}$} & \colhead{$f_{\mathrm{cluster}}$} & \colhead{$f_{\mathrm{galaxy}}$} & \colhead{$f_{\mathrm{emission}}$} & \colhead{$f_{\mathrm{cluster, W}}$}
}

\startdata
54 & 23.400054 & 30.709211 & 2.79 & 0.9667 & 0.2667 & 0.0000 & 0.7000 & 0.3235\\
57 & 23.469856 & 30.730419 & 3.84 & 0.9583 & 0.3583 & 0.0000 & 0.6000 & 0.4373\\
60 & 23.527783 & 30.815464 & 3.74 & 0.9667 & 0.2500 & 0.0000 & 0.7167 & 0.2552\\
114 & 23.491020 & 30.818007 & 2.37 & 0.9344 & 0.3771 & 0.0000 & 0.5574 & 0.4062\\
163 & 23.495319 & 30.819708 & 2.80 & 0.9344 & 0.6167 & 0.0000 & 0.3167 & 0.8384
\enddata
\tablecomments{Table \ref{tab:emissionreg} is published in its entirety in the electronic edition of the {\it Astrophysical Journal}.  A portion is shown here for guidance regarding its form and content.}
\end{deluxetable*}
\vspace{-2em}


\section{SLUG Integrated Light Fitting: Comparison to CMD-based Fits for M31 Clusters} \label{sec:app_slugM31}

To determine the reliability of SLUG integrated light fits, we use previous CMD-based cluster fits for M31 clusters observed by PHAT \citep{Johnson16_gamma} and compare to SLUG results for these clusters. Specifically, we identify a sample of clusters with good CMD fits and best fit log(Age/yr) $<$ 8.5. To compute the SLUG fits for the M31 clusters, we adopt the same fitting techniques we use for M33 that are described in Sec. \ref{sec:slugfitting}. We use of the same model grid for fitting, but we adopt a different assumed distance: 24.47 for M31. We report the SLUG fitting results for a sample of 885 M31 star clusters in Table \ref{tab:slug_m31}. 

We observe that a significant number of M31 SLUG fits have large quoted uncertainties. We consider a SLUG fit to have a large error if its 84th percentile minus 16th percentile value is $\ge$1.2 dex in age, or $\ge$1.3 dex in mass. This selection identified 122 clusters from the M31 sample with large SLUG uncertainties. We opt to flag these most uncertain fits in the results table and omit them from the CMD vs.\ SLUG comparison below.

\begin{figure*}
    \includegraphics[width=\textwidth]{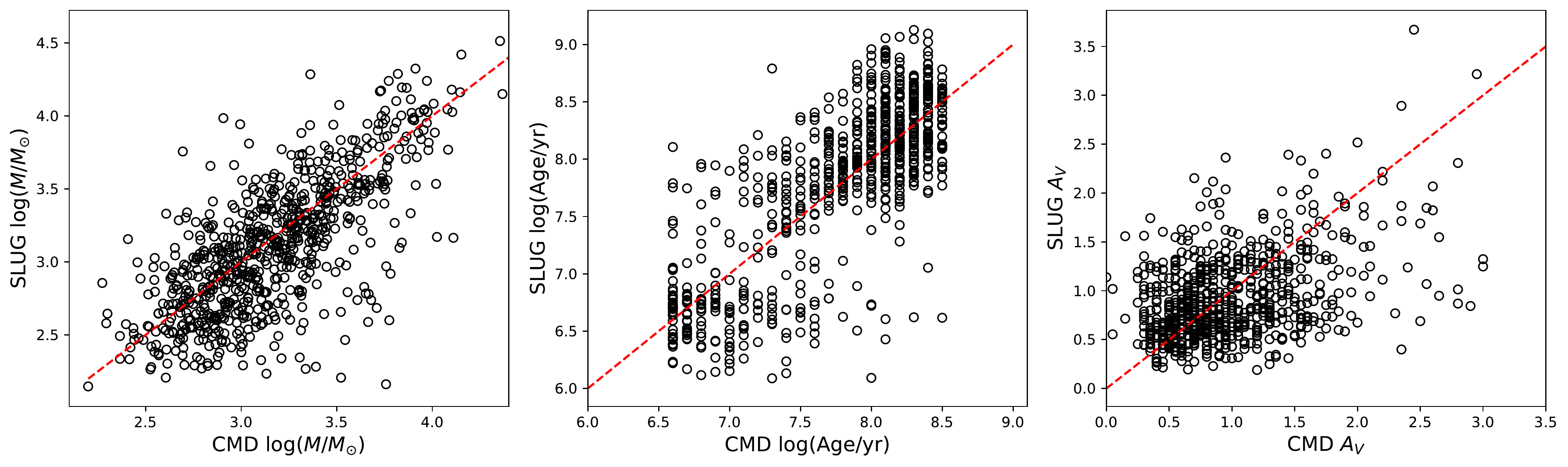}
    \caption{A comparison between SLUG integrated light fits and CMD-based fits for 763 clusters in M31.  Panels show distribution of results relative to the one-to-one relation (dashed red line) for cluster mass (left), age (middle), and $A_V$ (right).}
    \label{fig:SLUG_CMDcomp}
\end{figure*}

Figure~\ref{fig:SLUG_CMDcomp} compares results in mass, age, and $A_V$ from the cluster's CMD and SLUG fits. The reliability of integrated light SLUG values for mass is demonstrated on the top panel. The mass values for previous CMD estimates follow a 1:1 relationship with the new SLUG mass values. The same cannot be clearly stated for the SLUG age and $A_V$ values based on our comparison to CMD derived properties. 

The comparison of fitted ages reveals a gap in SLUG ages between log(Age/yr) of 7.0 and 7.5, whereas previous CMD ages are consistent with a continuous distribution for the last 300 Myr \citep{Johnson16_gamma}. The gap seems to be due to SLUG fitting clusters in this age range with either younger ($<$10$^7$ yrs) or older ($>$10$^7.5$ yrs) ages. Given this significant gap in the age distribution, similar to a known artifact in deterministic integrated light fitting at this same age that coincides with the age of emergence for evolved supergiant stars \citep[e.g., see][]{Fouesneau10}, we do not consider the SLUG ages to be as reliable as the CMD ages. At older ages ($>$10$^{8.5}$ yrs), however, SLUG ages become the sole option for age determination due to the difficulty of fitting CMDs in the case where the main sequence turnoff lies below the completeness limit for the resolve star photometry catalogs.

\pagebreak 

\begin{deluxetable*}{ccc|ccc|ccc|ccc}
\centering
\tabletypesize{\footnotesize}
\setlength{\tabcolsep}{0.05in}
\tablecaption{SLUG Results for PHAT M31 Clusters \label{tab:slug_m31}}
\tablehead{
\colhead{APID} & \colhead{Error Cut Flag} & \colhead{Filters Available} & \multicolumn{3}{|c|}{log($Mass/M_{\odot}$)} & \multicolumn{3}{c|}{log($Age/yr$)} & \multicolumn{3}{c}{$A_V$}  \\
\colhead{} & \colhead{} & \colhead{} & \multicolumn{1}{|c}{$SLUG_{16}$} & \multicolumn{1}{c}{$SLUG_{50}$} & \multicolumn{1}{c|}{$SLUG_{84}$} & \multicolumn{1}{c}{$SLUG_{16}$} & \multicolumn{1}{c}{$SLUG_{50}$} & \multicolumn{1}{c|}{$SLUG_{84}$} & \multicolumn{1}{c}{$SLUG_{16}$} & \multicolumn{1}{c}{$SLUG_{50}$} & \multicolumn{1}{c}{$SLUG_{84}$}\\
\colhead{} & \colhead{} & \colhead{} & \multicolumn{1}{|c}{$CMD_{16}$} & \multicolumn{1}{c}{$CMD_{50}$} & \multicolumn{1}{c|}{$CMD_{84}$} & \multicolumn{1}{c}{$CMD_{16}$} & \multicolumn{1}{c}{$CMD_{50}$} & \multicolumn{1}{c|}{$CMD_{84}$} & \multicolumn{1}{c}{$CMD_{16}$} & \multicolumn{1}{c}{$CMD_{50}$} & \multicolumn{1}{c}{$CMD_{84}$} 
}

\startdata
2 & T & 4 & 2.39 & 3.61 & 3.78 & 7.62 & 8.26 & 8.70 & 0.31 & 0.66 & 1.16 \\
& & & 3.95 & 3.98 & 3.98 & 8.4 & 8.4 & 8.4 & 1.1 & 1.15 & 1.15  \\
\hline
5 & F & 6 & 3.52 & 3.62 & 3.70 & 8.32 & 8.55 & 8.68 & 0.30 & 0.54 & 0.91 \\
& & & 3.40 & 3.41 & 3.45 & 8.3 & 8.4 & 8.4 & 0.44 & 0.45 & 0.65  \\
\hline
6 & F & 6 & 3.66 & 3.77 & 3.89 & 8.17 & 8.29 & 8.42 & 0.39 & 0.64 & 0.85 \\
& & & 3.81 & 3.83 & 3.83 & 8.4 & 8.5 & 8.5 & 0.5 & 0.5 & 0.65 \\
\hline
7 & F & 4 & 2.74 & 3.17 & 3.34 & 7.81 & 8.11 & 8.29 & 0.36 & 0.63 & 0.98 \\
& & & 3.12 & 3.16 & 3.16 & 7.9 & 8.2 & 8.2 & 0.4 & 0.45 & 0.55 \\
\hline
14 & F & 4 & 3.28 & 3.77 & 3.93 & 7.62 & 7.99 & 8.29 & 0.49 & 0.81 & 1.15 \\
& & & 4.08 & 4.08 & 4.13 & 8.2 & 8.2 & 8.3 & 1.15 & 1.25 & 1.25 
\enddata
\tablecomments{Table \ref{tab:slug_m31} is published in its entirety in the electronic edition of the {\it Astrophysical Journal}.  A portion is shown here for guidance regarding its form and content. The 16th, 50th, and 84th percentile values extracted for the mass, age, and $A_V$ PDFs from SLUG and CMD fitting results are reported. The filer count of the best available passband combination is listed, along with a boolean flag identifying cases of high uncertainty.}
\end{deluxetable*}


\bibliographystyle{aasjournal}
\bibliography{lcjlit}


\end{document}